\documentclass[10pt,sigplan,letterpaper,twocolumn]{article}
\usepackage[10pt,nocopyright]{sigmin}

\usepackage{listings}
\usepackage{color}
\usepackage{dingbat}
\usepackage[short]{optidef} 

\usepackage{epsfig,makecell}                                      
\usepackage[hyphens]{url}                                         
\usepackage[breaklinks,colorlinks]{hyperref}                      
\hypersetup{citecolor=blue,linkcolor=blue}                        
\usepackage{amsmath,amsopn,amssymb,amsthm}                        
\usepackage{thmtools}
\usepackage{thmbox}
\usepackage[toc,page]{appendix}
\usepackage{subcaption}                                           
\usepackage{endnotes,microtype,xspace,fancyvrb,multirow} 
\usepackage{epigraph} 
\usepackage{stfloats}

\usepackage{booktabs}      
\usepackage{tabularx}
\usepackage{array,underscore,relsize}                             
\usepackage[T1]{fontenc}                                          
\usepackage{times}                                                
\usepackage{fancyhdr,lastpage}                                    
\usepackage{enumitem}                                             
\usepackage[labelfont=bf,font=normal,skip=1pt]{caption}           
\usepackage[export]{adjustbox}                                    
\usepackage{breakurl}                                             
\usepackage{pifont}                                               
\usepackage{tablefootnote}                                        
\usepackage{bbding}                                        
\usepackage[table]{xcolor} 
\usepackage[linesnumbered,vlined,boxed,commentsnumbered,ruled]{algorithm2e}
\usepackage{float}
\DontPrintSemicolon
\newcommand{\ra}[1]{\renewcommand{\arraystretch}{#1}}
\newcommand{\nospacestitle}[1]{\noindent{\bf #1}}

\usepackage{amssymb}
\usepackage[normalem]{ulem}
\usepackage[hang,flushmargin]{footmisc}
\usepackage{textcomp}
\usepackage{graphicx}
\usepackage{authblk}

\usepackage[available]{usenixbadges}

\usepackage[compact]{titlesec}
\titleformat*{\section}{\large\bfseries}
\titleformat*{\subsection}{\normalsize\bfseries}
\titleformat*{\subsubsection}{\normalsize\bfseries}
\titlespacing{\section}{0pt}{3ex}{1ex}
\titlespacing{\subsection}{0pt}{2ex}{1ex}

\hypersetup{
  colorlinks,
  linkcolor={red!50!black},
  citecolor={blue!50!black},
  urlcolor={blue!80!black}
}

\newcommand{\sys}{{\textsc{BlitzScale}}} 
\newcommand{\maas}{\textsc{MaaS}}
\newcommand{\blitz}{\textsc{BlitzScale}}
\newcommand{\base}{ServerlessLLM}

\newcommand{\fig}[1]{Figure{~\ref{#1}}}

\newcommand{\one}{\texttt{\uppercase\expandafter{\romannumeral1}}}
\newcommand{\two}{\texttt{\uppercase\expandafter{\romannumeral2}}}

\newcommand{\DY}[1]{\textcolor{orange}{DY: #1}}
\newcommand{\TODO}[1]{\textcolor{red}{TODO: #1}}

\newcommand{\stitle}[1]{\vspace{1.1ex}\noindent{\bf #1}}

\usepackage{tikz}
\usepackage{colortbl}
\usetikzlibrary{tikzmark,calc}

\usepackage{balance}
\newcolumntype{P}[1]{>{\centering\arraybackslash}p{#1}}

\definecolor{appcolor}{RGB}{191,255,255}

\makeatletter
\renewcommand\AB@affilsepx{ \quad\protect\Affilfont \, } 
\makeatother

\SetKwInput{KwInput}{Input}
\SetKwInput{KwOutput}{Output}
\SetKwComment{Comment}{$\triangleright$\ }{}

\begin{document}

\title{\Large \bf{{\textsc{BlitzScale}}: Fast and Live Large Model Autoscaling with $O(1)$ Host Caching}}

\setlength{\affilsep}{0.5em}
\author[1]{Dingyan Zhang, Haotian Wang$^\dagger$, Yang Liu$^\dagger$, Xingda Wei\,{\Envelope}}
\author[2]{Yizhou Shan}
\author[1]{Rong Chen, Haibo Chen}

\affil[1]{\vspace{-2.mm}Institute of Parallel and Distributed Systems, Shanghai Jiao Tong University}
\affil[2]{Huawei Cloud\vspace{-1.mm}}

\date{}
\maketitle
\def\thefootnote{$\dagger$}\footnotetext{These authors contributed equally to this work.}
\def\thefootnote{\Envelope}\footnotetext{Xingda Wei is the corresponding author (\url{wxdwfc@sjtu.edu.cn}).}
\def\thefootnote{}\footnotetext{\textit{Proceedings of the 19\textsuperscript{th} OSDI Conference}, Boston, MA, USA, 2025.}

\renewcommand{\thefootnote}{\arabic{footnote}}

\frenchspacing

\begin{abstract}
    \noindent
    Model autoscaling is the key mechanism to achieve serverless model-as-a-service, 
    but it faces a fundamental trade-off between scaling speed and storage/memory usage to cache parameters, 
    and cannot meet frequent scaling requirements across multiple hosts. 
    The key problem is that data plane performance is slow, 
    and scaled instances remain stopped while parameters are loading.     

    In this paper, we first show that the data plane
    can be made \emph{fast} with no or $O(1)$ caching
    by loading parameters through the compute network between GPUs because:
    (1) its speed is comparable to host cache and is underutilized, and
    (2) scaling multiple instances requires no or $O(1)$ caching with network-optimized multicast.
    Second, autoscaling can be made \emph{live} by
    breaking the scaling abstraction for inference
    from a coarse-grained instance-level to a fine-grained layer-level.
    This allows us to offload the layer computation from the overloaded serving instances
    to the scaled ones without waiting for the parameters to be fully loaded.

    Under real-world workloads,
    our system {\sys} achieves up to 94\,\% lower tail latency reductions compared to
    state-of-the-art autoscaling system (ServerlessLLM),
    and it reduces the GPU time used for serving by 49\,\% when compared with
    serving systems that do not support autoscaling like DistServe and vLLM with the same
    service-level-agreement.
\end{abstract}

\section{Introduction}
\label{sec:intro}

\noindent
Recent years have seen rapid growth in applications powered
by deep learning models like large language models (LLMs)~\cite{aws-rekognition,chatgpt,copilot,sora,stability-ai}.
Due to the huge computation requirements,
these models are typically served in model-serving-as-a-service systems ({\maas})~\cite{serverless-together-ai,serverless-ray,deepinfra,DBLP:conf/asplos/YangZLZLZCL22,ali-serverless-llm,aws-serverless-llm,DBLP:journals/corr/abs-2401-14351},
which manage a cluster of accelerators (e.g., GPUs) 
and provision an appropriate number of serving \emph{instances} containing GPUs
to each model deployed.

An {\maas} system has two design objectives:
\emph{maximizing goodput}---the number of requests
that meet the service level objective (SLO), and
\emph{minimizing instances provisioned} to each model to improve hardware utilization.
Achieving both is challenging
 due to the unpredictable short-term fluctuations in a model's instance demands,
 (5\,$\times$ required within 2 seconds),
 because the request arrival rate bursts at seconds-level~\cite{alpaserve,DBLP:conf/nsdi/0025TKS23},
 where the memory usage of each request is also unpredictable due to the
 auto-regressive nature of LLMs~\cite{alpaserve,DBLP:conf/nsdi/0025TKS23,DBLP:journals/corr/abs-2401-14351} (see also {\fig{fig:problem-statement} and \textsection{\ref{sec:problem-statement}}}).

 Model autoscaling is a promising solution~\cite{DBLP:conf/asplos/YangZLZLZCL22,serverless-ray,DBLP:journals/corr/abs-2401-14351,kserve}.
 With autoscaling,
  a {\maas} system only provisions the average number of instances required
  over the long term for each served model, which remains relatively stable~\cite{wang2024burstgpt}.
  This improves utilization.
  Upon bursts,
  the system automatically scales new instances,
  avoiding SLO violations due to request queueing caused by insufficient instances provisioned.

Autoscaling speed is critical in minimizing SLO violations
because the queued requests are not served until the instances are scaled.
For instance, the inference time of a Llama3-8B is 80-900\,ms on commodity GPU (A800),
while users expect a tight response time (< 1 second) for scenarios like chatbot~\cite{latency-study,DBLP:conf/osdi/ZhongLCHZL0024,aws-latency}.
Meeting such tight SLO requires less than 500\,ms scaling time,
but achieving this is challenging especially for LLMs
with 10-400\,GB parameters.
The key reason is
the \textbf{slow data plane} of autoscaling---the process of loading the model parameters to instances' GPUs.
While high bandwidth SSDs
are utilized in current work~\cite{DBLP:journals/corr/abs-2401-14351},
the speed provided by SSDs of GPU servers (2-10\,Gbps per GPU~\cite{googleinstance,awsinstance,volcanoinstance})
is still far from ideal.
For instance, loading Llama3-8\,B to a GPU takes 12.8 seconds with 10\,Gbps SSD.
Another factor obstructing fast scaling is that existing scaling methods are
\textbf{stop-the-world}:
the scaled instances cannot serve requests until all parameters are loaded.
This implies that autoscaling is directly bottlenecked by the data plane.

To mitigate the above issues, state-of-the-art systems like ServerlessLLM
further adopt a multi-tiered caching system
by caching model parameters in the host (CPU) DRAM to accelerate the data plane~\cite{DBLP:conf/osdi/GujaratiKAHKVM20,DBLP:conf/eurosys/JeongBA23}.
Under cache hit, they can leverage the fast CPU-GPU link (e.g., 256\,Gbps PCIe) to load parameters.
However, achieving a high hit rate is unfeasible:
ServerlessLLM reports a hit rate of 40--75\,\%,
which is confirmed by us (see \textsection{\ref{sec:bg-motiv}}).
The root cause is that a {\maas} typically hosts many models,
thus achieving a 100\,\% cache hit requires caching all these models on the DRAM of each host,
clearly impractical.
Vendors typically host many models because
there are hundreds of popular open-source model families designed for different purposes~\cite{huggingface}.
Meanwhile,
each model family has different scales for balancing the serving cost and accuracy~\cite{qwens}.
Finally,
developers can upload their customized fine-tuned models based on open-source models~\cite{aws-serverless-llm}.

To achieve fast model scaling without relying on cache hit,
we make the following two key contributions:

\stitle{1. Data plane made fast with $O(1)$ or no caching with compute network multicast.\,}
First, a {\maas} is backed by fast GPU-GPU/CPU compute fabrics~\cite{nvidia-compute-network},
which are 100--400\,Gbps RDMA and even 16\,Tbps NVLink~\cite{googleinstance,awsinstance,volcanoinstance}---much
faster than SSD and comparable (or even faster than) CPU-GPU PCIe.
The compute fabric is used for data transfer during serving
and we found it largely under-utilized,
i.e., up to 7.4\,\% of total bandwidth
even in network-heavy workloads like serving LLMs with prefill and decode disaggregation~\cite{DBLP:conf/isca/PatelCZSGMB24, DBLP:conf/osdi/ZhongLCHZL0024,DBLP:journals/corr/abs-2401-11181,DBLP:conf/ppopp/CaiLMMMNS21,DBLP:journals/corr/abs-2404-09526} (\textsection{\ref{sec:bg-motiv}}).
Thus, we can borrow such fast links for accelerating the data plane of autoscaling.

Second, network-based data plane requires no or minimal caching to achieve fast scaling. 
Specifically,
if a model is already deployed on some instances,
we can directly multicast the parameters from deployed instances through the network, 
eliminating the need for caching.
Such multicast is extremely efficient
because a serial forwarding multicast~\cite{DBLP:conf/icpp/VerstoepLB96}
can load bulk data (e.g., model parameters), regardless of the number of receivers. 
Even if no instance is deployed, multicast can be done with $O(1)$ host caching by
simply broadcasting the parameters from the host with the cached model. 
This $O(1)$ caching per-model allows us to avoid all cache misses
since the aggregated host memory of all machines is sufficient to cache all models served by a {\maas}. 

Although fast networking can significantly accelerate the data plane
with minimal caching,
a stop-the-world loading remains a bottleneck in cases
when the networking is not fast enough.
For example,
to achieve at most 40\,\% SLO violations when serving a BurstGPT workload with Qwen2.5-72B,
the system needs to achieve a tight 500\,ms stop time.
Achieving so requires 576\,Gbps per-GPU\footnote{\footnotesize{72\,B model requires at least four GPUs per-instance for serving.}}
parameter transfer bandwidth,
far exceeding
the available bandwidth of typical compute network setups (e.g., 200\,Gbps per-GPU)
and even when caching at the host (256\,Gbps PCIe).
Thus, we argue that an ideal parameter loading should be \emph{live}:
before the data loading finishes, the scaled instance should be able to serve requests.

\stitle{2. Data plane made live with fine-grained scaling abstraction and cooperative execution. \,}
Model scaling cannot be live using traditional
on-demand data loading techniques commonly found in serverless computing~\cite{mitosis,DBLP:conf/eurosys/WangHW19,DBLP:conf/sosp/HuangZMLLCJLSZL24}
or inference loading overlap in PipeSwitch~\cite{pipeswitch} (\textsection{\ref{sec:overview}}),
because an instance can only emit results once all the parameters are loaded.
This stop is rooted in the coarse-grained scaling abstraction of existing systems:
they can only scale and serve at the instance level.
To realize live scaling,
our key insight is that
\emph{models can be served in a fine-grained, layer-by-layer manner}.
With this fine-grained layer-wise scaling,
we can offload part of the layer's computation from overloaded instances to scaled instances
with cooperative execution,
thus improving the overall serving throughput
even before the scaled instance has loaded all the parameters.

\stitle{Challenges and solutions.\,}
First,
utilizing network-based multicast is non-trivial in our setup.
Though the mechanism of multicast is simple, i.e.,
simply forwarding parameters between instances with the network,
the challenges lie in generating the multicast plan,
i.e., determining how the data flows between instances.
First, we need to quickly generate an efficient plan online on
diverse network topologies since our sources and destinations are dynamically determined,
but generating an optimal plan is NP-hard on
heterogeneous networks in serving clusters~\cite{DBLP:journals/jpdc/BhatRP03}.
Second, we need to avoid network interference between the scaling and serving,
otherwise, we observed a 1.5\,$\times$ longer scale time and 50\% degraded tail TBT (\textsection{\ref{sec:overview}}).
Current solutions~\cite{DBLP:conf/asplos/CowanMMSX23,DBLP:conf/ppopp/CaiLMMMNS21,DBLP:conf/sc/GhazimirsaeedZR20} mainly
target offline scenarios like training,
so they can tolerate long plan generation time and
don't need to consider interference from serving workloads.
To address the issue, we propose a
model-aware multicast planner,
which leverages the key features of compute network and the
static data flow in model serving to quickly generate a near-optimal,
interference-free multicast plan for scaling (\textsection{\ref{sec:design-plan}}).

Second, it is challenging to schedule how requests are executed
between deployed and live scaling instances,
i.e., which instance executes which layers.
The challenge lies in the fact that the serving capability of the scaling instances
is limited---it can only execute layers with parameters loaded,
and this capability is dynamically changing.
A naive best-effort scaling that executes as many layers as possible
cannot balance the load because at the beginning of autoscaling,
the new instances can only execute few layers, 
with requests still queued at the overloaded instances.
A better solution is to adjust the load holistically by considering future incoming layers,
and we realize this with a ZigZag pipeline scheduling
and achieve 50\% tail latency reduction under bursty workloads
(\textsection{\ref{sec:design-pp}}).

\stitle{Demonstration with {\sys}. \,}
We built {\sys}, an {\maas} system with the fastest autoscaling speed with $O(1)$ caching.
We adopted a global parameter pool to cache the model parameters across all the machines,
and integrated the aforementioned interference-free multicast plan for scaling
and efficient ZigZag scheduling-based live scheduling.
To show the effectiveness of {\sys}, 
we evaluated {\sys} by running real-world traces
(i.e., BurstGPT~\cite{wang2024burstgpt}, AzureCode and AzureConv~\cite{AzurePublicDataset})
across a variety of recent models
with different sizes and architectures, including Llama3-8B, Mistral-24B, and Qwen2.5-72B.
First, {\sys} has 47--75\,\% shorter time-to-first token,
and has up to 94\,\% shorter time-between-tokens 
than the state-of-the-art work (ServerlessLLM~\cite{DBLP:journals/corr/abs-2401-14351}). 
Second, compared to serving systems without autoscaling support,
i.e., vLLM~\cite{vllm} and DistServe~\cite{DBLP:conf/osdi/ZhongLCHZL0024},
{\sys} reduces the GPU used for serving a single model by 49\,\%
with no SLO violations compared to an over-provisioning setup
that provisions the GPUs based on the maximum request rate.
{\sys} is open-sourced at \burl{https://github.com/blitz-serving/blitz-scale}. 

\section{Background: {\maas} and Autoscaling}
\label{sec:bg}

\subsection{System setup: model-serving-as-a-service (\maas)}
\label{sec:model}     

\begin{figure*}[!t]
    \centering
    \includegraphics[width=1.01\textwidth,center]{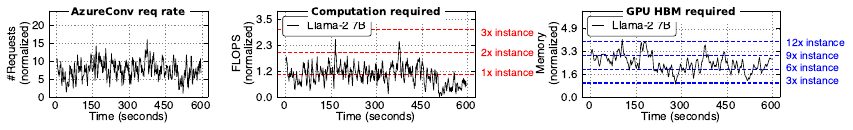} \\[0pt]
    \begin{minipage}{1\linewidth}
    \caption{\small{\textit{
        The timeline of request incoming rate of a real-world AzureConv~\cite{AzurePublicDataset} trace (a), 
        its computation (b) and memory requirements 
        (c) when serving this workload without SLO violation. 
    }}}
    \label{fig:problem-statement}
    \end{minipage} \\[-15pt]
\end{figure*} 

\begin{figure}[!t]
        \begin{minipage}{1\linewidth}
        \hspace{-1mm}
        \centering    
        \includegraphics[width=\columnwidth, trim=0.25cm 13.8cm 12cm 0.25cm, clip]{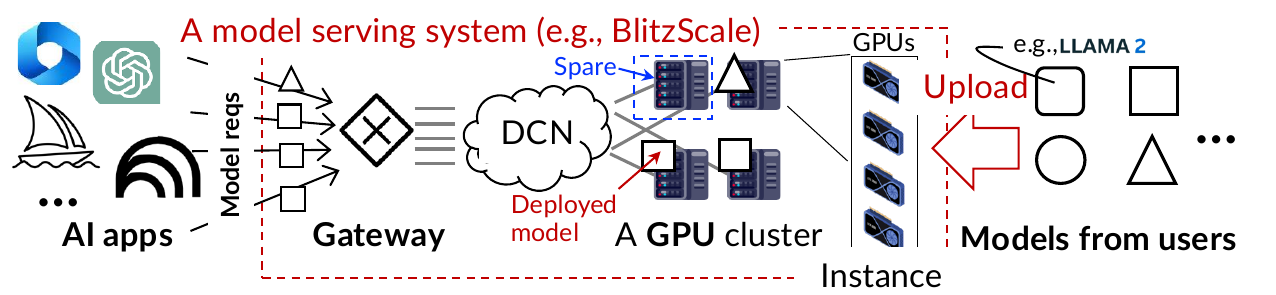} \\[1pt]
        \end{minipage} \\[3pt]
        \begin{minipage}{1\linewidth}
        \caption{\small{\textit{
            An illustration of Model as a Service ({\maas}) system. 
            DCN states for \uline{d}ata \uline{c}enter \uline{n}etwork. 
        }}}
        \label{fig:setup}
        \end{minipage} \\[-20pt]
        \end{figure} 

\nospacestitle{{\maas} system setup and the serving instance.\,}
{\sys} targets a {\maas} scenario~\cite{serverless-together-ai,serverless-ray,deepinfra,DBLP:conf/asplos/YangZLZLZCL22,ali-serverless-llm,aws-serverless-llm,DBLP:journals/corr/abs-2401-14351}:
the cloud allows the users to deploy their model serving on its managed hardware,
and only charges users based on the number of requests processed within SLO~\cite{aws-serverless-llm,ali-serverless-llm}.
The model can be popular open-source models like Qwen~\cite{qwens}
or customized models uploaded by the users.
Thanks to the above pricing strategy,
the cloud can dynamically adjust hardware resources to a specific model
to maximize its hardware utilization.
{\fig{fig:setup}} shows a typical deployment of a {\maas} system.
For each user-deployed model,
the system dynamically allocates the required hardware resources (GPUs) to serve the
inference requests of the model.
Note that due to the diversity of AI applications,
there would be hundreds or even thousands of models served simultaneously by a {\maas} system~\cite{alibailian}.

In this paper,
we use \emph{instance} to denote a set of GPUs storing
a complete copy of a model parameter for serving this model.
An instance can have a single GPU or multiple GPUs when the parameter
is large and sharded across them, e.g., with tensor parallelism~\cite{narayanan2021efficient}.
Because each instance has a maximal serving throughput,
the cloud can deploy multiply instances of the same model
by provisioning multiple sets of GPUs, 
where the number of instances is dynamically scaled based on the incoming request rate.
{\sys} supports autoscaling with all existing model serving methods at the instance level. 

\stitle{Serving within an instance: non-LLM \& LLM. \,}
Each serving instance processes requests in the following workflow:
it queries the model in a layer-by-layer computation paradigm (see {\fig{fig:overview}} (a))
and gets the final results.
For simple models like vision models,
the model is queried once with the input data (image).
On the other hand, for large language models (LLMs)~\cite{DBLP:conf/nips/VaswaniSPUJGKP17},
the request queries the model multiple times:
the model is first queried with input text (prompt)
to produce a result (token).
This first query is typically termed \emph{prefill}.
The token is then used to generate subsequent tokens iteratively
until the model returns an end-of-sequence token.
The auto-regressive phase is termed \emph{decode}.

We note two important features of LLMs.
First,
the performance for prefill and decode is measured separately.
Prefill is evaluated with the time-to-first-token (TTFT)
while decode is evaluated using time-between-tokens (TBT).
Second, the LLM query is stateful:
the intermediate results---usually termed as \emph{KVCache}---are cached in GPU memory
during the auto-regressive phase of a request for acceleration.

\stitle{Serving across instances:  prefill and decode (PD) disaggregated LLM serving. \,}
Observing the different computing paradigms of prefill and decode,
recent works propose separating the instances for prefill and decode (PD disaggregation)
when processing serving requests~\cite{DBLP:conf/isca/PatelCZSGMB24, DBLP:conf/osdi/ZhongLCHZL0024,DBLP:journals/corr/abs-2401-11181}.
Specifically, for each request, one instance processes the prefill phase (prefill instance )
and another instance (decode instance) processes the decode phase.
This paradigm requires excessive data movement between the two instances
because the prefill instance needs to transfer the KVCache to the decode instance.
{\sys} works for both PD disaggregated and non-disaggregated LLM serving.

\subsection{Dynamic hardware demands when serving a model}
\label{sec:problem-statement}

\noindent
Determining the hardware requirements, i.e.,
the right number of instances for serving a model is challenging
because the hardware demands are unpredictable and fluctuating.
First,
the incoming request rate for a serving workload fluctuates over time
and is hard to predict~\cite{DBLP:journals/corr/abs-2401-14351,DBLP:conf/usenix/ShahradFGCBCLTR20,DBLP:conf/nsdi/0025TKS23}.
{\fig{fig:problem-statement}} (a) presents the number of requests sent to a single model service
over time from a real-world trace---BurstGPT~\cite{wang2024burstgpt}:
the incoming inference requests increase 5\,$\times$ within 2 seconds with no predictable trend.
Since the FLOPS of an instance is fixed, 
the unpredictable rate causes 
the computation requirement---FLOPS required to finish the pending
requests within SLO---unpredictable.
{\fig{fig:problem-statement}} (b) confirms this by measuring the requirement 
of the prefill instances when serving the BurstGPT with Llama2-7B.

Second,
serving modern models like LLM
has non-trivial and unpredictable memory requirements.
As shown in
{\fig{fig:problem-statement}} (c),
the KVCache usage of the decode instances is multiple times larger than the memory capacity
of a single instance and fluctuates over time (3--12\,$\times$) when serving the BurstGPT workload with Llama2-7B.
The root cause is that
the KVCache of requests must be stationary in GPU memory during the decode phase.
The KVCache of requests are large, e.g., 190--760\,GB for Llama2-7B to serve BurstGPT, 
and the stay time is unpredictable due to the auto-regressive nature of LLMs.
To avoid performance losses when out of memory,
a {\maas} system must provision sufficient instances
to hold KVCache from ongoing requests,
so the number of instances required by a model also unpredictably fluctuates.

\subsection{Model autoscaling for handling dynamic demands}

\noindent
Model autoscaling, which dynamically deploys\footnote{\footnotesize{It also stops a serving instance to scale down.
Since scaling down is simpler, we omit its details for brevity.}}
serving instances on spare GPUs to scale up the serving capability, is
a promising solution to handle fluctuated and unpredictable computation and memory demands~\cite{DBLP:journals/corr/abs-2401-14351,DBLP:conf/asplos/YangZLZLZCL22,serverless-ray}.
The rationale is that though a single model's workload is unpredictable,
the aggregated workloads of all models served by a platform are relatively stable~\cite{DBLP:conf/nsdi/0025TKS23}.
Thus, when the load of a specific model service increases,
we are able to find spare GPUs from other models to scale up the serving capability of this model. 

Autoscaling an instance requires two basic steps:
(1) initialize a proper execution context, e.g., create CUDA contexts (control plane)
and 
(2) load the model parameters to the GPUs' memory (data plane). 
We focus on (2) because (1) can be minimized with recent advances
in GPU startup methods like checkpoint
and restore~\cite{DBLP:journals/corr/abs-2405-12079,DBLP:conf/asplos/ZengXGCL25} 
and our Rust/C++-based serving platform
(see also \textsection{\ref{sec:eval-ablation}}).
For (2), the state-of-the-art system---ServerlessLLM~\cite{DBLP:journals/corr/abs-2401-14351}
optimizes the data plane with SSD-optimized parameter loading.
Unfortunately, it does not account for the scaling speed required by models.
Our measurements in the next section
show that SSD-based scaling significantly lags behind applications' requirements.

\begin{figure*}[!t]
    \centering
    \includegraphics[width=1.01\textwidth,center]{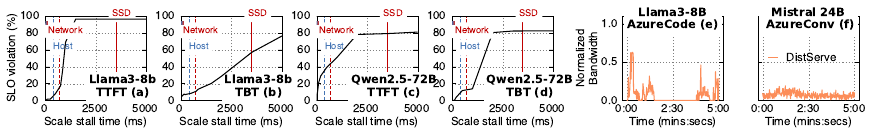} \\[2pt]
    \begin{minipage}{1\linewidth}
    \caption{\small{\textit{
        A characterization of SLO attainment for different inference cases (a)--(d)
        with varied duration of autoscaling stops on BurstGPT~\cite{wang2024burstgpt}. 
        (e) and (f): an analysis of compute network usage in serving workloads.
        The evaluation setup is in \textsection{\ref{sec:eval}}.
    }}}
    \label{fig:motiv-scale-speed}
    \end{minipage} \\[-5pt]
\end{figure*}

\section{Characterizing Scaling Requirements and Compute Network between Instances}
\label{sec:bg-motiv}  

\nospacestitle{Model autoscaling requires fast data plane. \,}
If the data plane speed is not fast enough, during burst period, the requests
still violate the SLO due to increased queueing time.
Specifically, SLO defines the tolerable end-to-end latency measured from the 
time a request is sent to the system to the time the inference response is returned.
Thus, the latency includes the queueing delays waiting for the scaled instance 
to be ready for inference.

To characterize how different scaling speeds affect SLO attainments,
we implemented a simulator on DistServe~\cite{DBLP:conf/osdi/ZhongLCHZL0024}
that provisions models to all GPUs and applies manual delays
based on the simulated speed for modeling different scaling speeds.
We set TTFT and TBT SLO based on inference speed of different models following
prior works~\cite{DBLP:conf/osdi/ZhongLCHZL0024}.
Specifically, we use 450\,ms and 150\,ms for Llama3-8B model,
and 1250\,ms and 200\,ms for Qwen2.5-72B model with tensor parallelism degree of 4.
\textsection{\ref{sec:eval}} describes the detailed evaluation setup.

{\fig{fig:motiv-scale-speed}} (a)--(d) shows the results: 
We can see that 
for a 72\,B model,
maintaining SLO violations below 60\,\%
requires a minimum per-instance scaling speed of 220\,Gbps per-GPU\footnote{\footnotesize{72\,B model uses four GPUs per-instance.}},
which is only achievable when the model parameters are loaded from the host memory.
The scaling time requirement correlates directly with inference time---our evaluated workload (BurstGPT~\cite{wang2024burstgpt})
has an average TTFT of 771\,ms (with queueing time).
Thus, to achieve 1250\,ms SLO for all requests,
the scale time must be below 500\,ms,
so a 576\,Gbps per-GPU network speed
is required (measured by dividing the parameter size by the scale time).
This far exceeds what vendor-provided per-GPU
SSDs bandwidth can deliver (2--10\,Gbps per-GPU~\cite{googleinstance,awsinstance,volcanoinstance}, 
detailed in \textsection{\ref{sec:appendix-hardware}}).

\begin{figure}[!t]
    \centering
    \includegraphics[width=1.01\linewidth,center]{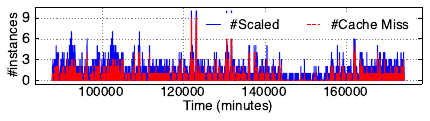} \\[1pt]
    \begin{minipage}{1\linewidth}
    \caption{\textit{\small{
        An analysis of host cache misses when running ServerlessLLM~\cite{DBLP:journals/corr/abs-2401-14351}
        on BurstGPT~\cite{wang2024burstgpt}.
    }}}
    \label{fig:motiv-cache-miss}
    \end{minipage} \\[-5pt]
\end{figure} 

\stitle{Loading model parameters from host memory is not effective due to misses. \,}
While caching the parameters on the host CPU memory 
can meet the scaling speed requirement for some setups (e.g., 8\,B, 24\,B)
with the fast host-GPU interconnects (256\,Gbps PCIe 4.0),
cache misses are common in real-world traces,
because the scarce host memory cannot support the caching all models deployed on the {\maas}. 
{\fig{fig:motiv-cache-miss}} presents the number of instances scaled and cache misses 
encountered in the BurstGPT 
workload using ServerlessLLM~\cite{DBLP:journals/corr/abs-2401-14351}. 
Following its setup,
we set a 5-minute keep-alive interval for caching models at the host. 
The miss rates range from 20--46\%, 
depending on the time, 
which aligns with the numbers reported in ServerlessLLM's paper (25--60\%).  
Interestingly, many misses occur when scaling multiple instances, 
because involving more hosts increases the probability of scaling 
a model on a host without the cached parameters.
Therefore, we still need accelerating scaling speed when the model parameters
are not cached at the host.

\begin{figure}[!t]
        \begin{minipage}{1\linewidth}
        \hspace{-1mm}
        \centering    
        \includegraphics[width=.99\textwidth, trim=0.25cm 11.9cm 14.4cm 0.25cm, clip]{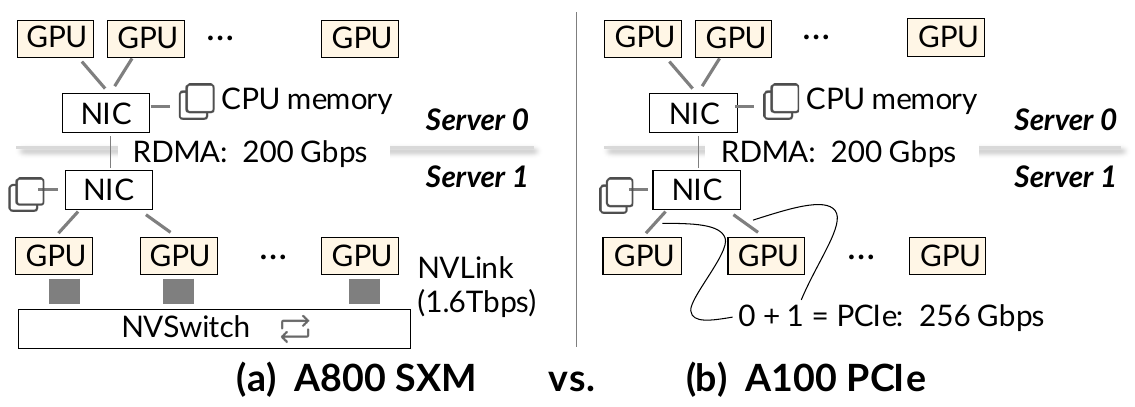} \\[1pt]
        \end{minipage} \\[4pt]
        \begin{minipage}{1\linewidth}
            \caption{\textit{\small{
            A illustration of networking in {\maas} with NVLink (a) and without it. 
            Note that the 256\,Gbps PCIe is shared between two GPUs attached to it~\cite{nvidia-all-to-all}.
        }}}
        \label{fig:gputopo}
        \end{minipage} \\[-10pt]
        \end{figure}

\stitle{Opportunity: fast and underutilized compute network. \,}
First, compute networks  between GPUs (and CPUs) have comparable or even faster speeds
than host-to-GPU link. As shown in {\fig{fig:gputopo}},
the inter-GPU network (RDMA) operates at 200\,Gbps,
which is close to the host-to-GPU PCIe speed (256\,Gbps). 
With NVLink, the speed is much faster. 
More importantly, these networks are underutilized during serving.
{\fig{fig:motiv-scale-speed}} (e) and (f) measure the peak
network usage of DistServe~\cite{DBLP:conf/osdi/ZhongLCHZL0024},
a PD disaggregated serving system that heavily utilizes the network due to KVCache transfers.
To measure peak usage, we provisioned all the GPUs for serving,
and evaluate a workload with the maximal request rate that our clusters can serve.
Even under peak load, more than 40\% of the network capacity is free,
opening up the opportunity to use the compute network for the scaling data plane.

\section{System Overview of {\sys}}
\label{sec:overview}

\definecolor{boxcolor}{HTML}{FFFFFF}
\definecolor{boxcolortwo}{HTML}{FFF6E6}

\newcommand{\fboxone}[1][1em]{%
  \tikz\filldraw[fill=boxcolor,draw=black] (0,0) rectangle (#1,#1);
}

\newcommand{\fboxtwo}[1][1em]{%
  \tikz\filldraw[fill=boxcolortwo,draw=black] (0,0) rectangle (#1,#1);
}

\begin{figure}[!t]
        \begin{minipage}{1\linewidth}
        \hspace{-1mm}
        \centering    
        \includegraphics[width=.99\textwidth, trim=0.25cm 11.6cm 22.3cm 0.25cm, clip]{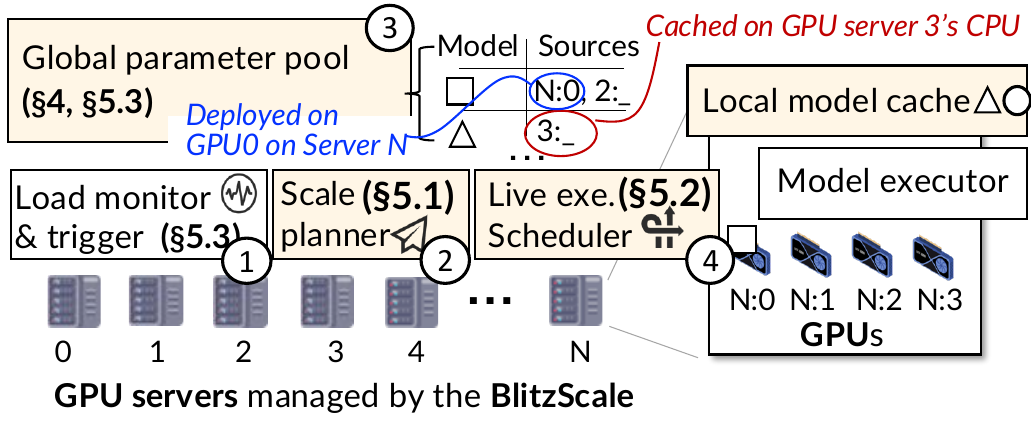} \\[1pt]
        \end{minipage} \\[3pt]
        \begin{minipage}{1\linewidth}
        \caption{\small{\textit{
            System architecture of {\sys}.
            New modules introduced by {\sys} are marked with {\fboxtwo}. 
        }}}
        \label{fig:arch}
        \end{minipage} \\[-10pt]
        \end{figure} 

\noindent
{\sys} scales models through the compute network 
to accelerate scaling even under cache misses on the host.
We achieve this by first managing model parameters---scattered across GPUs behind serving instances
 (for deployed models) 
and CPUs (cached at the host)---through a global parameter manager. 
The manager maintains a mapping between models and their sources. 
With the manager, we can quickly read parameters from these sources with the fast 
RDMA or NVLink.
Besides, we also offload computation from overloaded instances to 
instances with partially loaded parameters to achieve live scaling.        

\stitle{System architecture and workflow. \,}
{\fig{fig:arch}} shows our system architecture.
Like prior work~\cite{DBLP:conf/usenix/Romero0YK21,DBLP:journals/corr/abs-2401-14351,DBLP:conf/osdi/SunHZXZL024},
we have a load monitor (\ding{192})
tracking the serving load for each model service,
and deciding whether to scale and how many new instances are required (\textsection{\ref{sec:design-others}}).
On each machine, we further adopt off-the-shelf GPU kernels FlashInfer~\cite{flashinfer}
to query the model efficiently.
The key differences are twofold.
First, our scale planner (\ding{193}) will
derive a scaling plan that guides how to load parameters onto the scaled instances (\textsection{\ref{sec:design-plan}})
with compute network efficiently.
The planner consults the global parameter manager (\ding{194}) to identify the sources of model parameters.
In our example, the new instance can load the parameters of model {$\square$}
from the GPU0 on host N ($N,0$), or from host 2's CPU memory ($2,\_$).
Second,
during scaling,
our live execution (exe.) scheduler (\ding{195}) will redirect requests between instances to
fully utilize the scaled instances even before the parameters are fully loaded (\textsection{\ref{sec:design-pp}}).

\begin{figure}[!t]
        \begin{minipage}{1\linewidth}
        \hspace{-1mm}
        \centering    
        \includegraphics[width=.99\columnwidth, trim=0.25cm 11.3cm 17.7cm 0.25cm, clip]{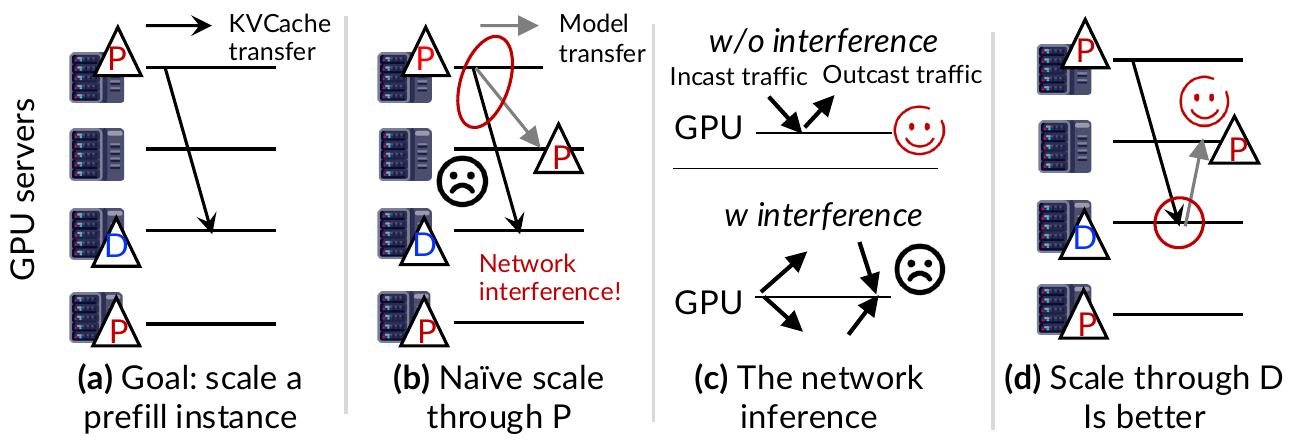} \\[1pt]
        \end{minipage} \\[5pt]
        \begin{minipage}{1\linewidth}
        \caption{\small{\textit{
            (a) An illustration of scaling a prefill instance for LLM PD disaggregated serving. 
            (b) Naively scaling from a prefill instance imposes network interference.
            (c) Interference can be avoided by leveraging the bi-directional feature of modern DCN networking. 
            (d) An improved scale plan with the bi-directional in mind. 
        }}}
        \label{fig:pd-scale}
        \end{minipage} \\[-10pt]
        \end{figure}    
        
\begin{figure}[!t]
    \centering
    \includegraphics[left, scale=1.1]{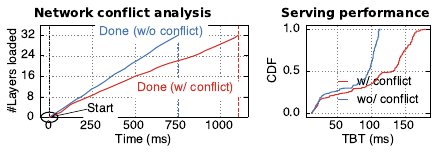} \\[0pt]
    \begin{minipage}{1\linewidth}
            \caption{\small{\textit{
            A characterization of network interference on (a) scaling speed and (b) serving performance. 
        }}}
    \label{fig:motiv-network-conflict}
\end{minipage} \\[-5pt]
\end{figure}

\stitle{Challenges and approaches. \,}
Despite leveraging fast networking, making autoscaling
fast and live needs to address the following challenges.

\emph{\uline{C\#1. Online interference-free scale plan generation.} \,}
Generating the scale plan is similar to generating a \emph{multicast} plan~\cite{DBLP:journals/jpdc/BhatRP03,DBLP:conf/asplos/CowanMMSX23,DBLP:conf/sc/GhazimirsaeedZR20,DBLP:conf/icpp/BanikazemiMP98,DBLP:conf/dsn/BehrensJBT18},
i.e., how to quickly distribute data (parameters) from some sources to targets.
There are two additional requirements for autoscaling.
First, the plan must be generated online on dynamically changing sources and targets,
but optimal plan generation is NP-hard~\cite{DBLP:journals/jpdc/BhatRP03} on a heterogeneous network.
Second, the plan needs to eliminate interference between loading and the serving workload.
{\fig{fig:pd-scale}} shows an example when a model is served via PD disaggregation.
In PD disaggregation serving workload KVCache is migrated from prefill to decode instances (a), 
and this migration overhead can be hidden~\cite{DBLP:conf/isca/PatelCZSGMB24}.
However, suppose we want to scale a prefill instance:
if we naively select a prefill instance as the source (b),
the scaling will compete the network bandwidth with the serving workload,
leading to 1.5\,$\times$ longer scale time as well as 50\% tail TBT increase due to the 
amplified KVCache migration overhead
({\fig{fig:motiv-network-conflict}} (b)).

To this end,
we design a serving-guided greedy plan generation method
based on three observations (\textsection{\ref{sec:design-plan}}).
First,
the network heterogeneity mainly comes from NVLink,
whose speed is extremely fast,
i.e., it can broadcast a Llama3\,8B to 8 GPUs within 120\,ms.
Thus, we can abstract instances linked with NVLink as a logical instance group
to eliminate NVLink from the network topology.
Second, loading parameters from the network is bandwidth-intensive,
so we can greedily construct serial forwarding chains~\cite{DBLP:conf/icpp/VerstoepLB96}
for multicast, which is optimal in the common case.
Finally, the network (RDMA) between GPU servers is bi-directional~\cite{288657,rdma-bidirectional},
meaning that the network flows of incast and outcast don't interfere (c).
Thus,
we can leverage this feature to avoid interference
by removing flows in the same direction on the same network link during plan generation.
For example, we can load the parameters from the decode instance to the prefill instance (see {\fig{fig:pd-scale}} (d)).

\begin{figure*}[!t]
        \begin{minipage}{1\linewidth}
        \hspace{-1mm}
        \centering    
        \includegraphics[width=.95\columnwidth, trim=0.25cm 10.38cm 2.5cm 0.25cm, clip]{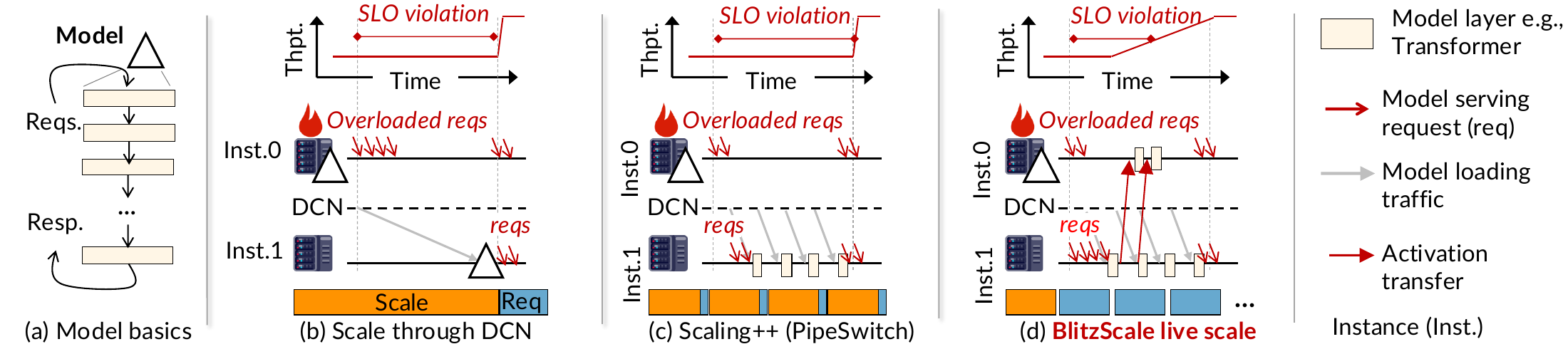} 
        \end{minipage} \\[8pt]
        \begin{minipage}{1\linewidth}
        \caption{\small{\textit{
            (a) An overview of how model executes requests.
            (b) An illustration of naive model scale through DCN.
            (c) An illustration of an optimized scaling method with overlapped execution~\cite{pipeswitch},
            but it still cannot be live.
            (d) How {\sys} scales models lively.
        }}}
        \label{fig:overview}
        \end{minipage} \\[-10pt]
        \end{figure*} 

\vspace{1.ex}
\emph{\uline{C\#2. Realizing live autoscale.} \,}
Live autoscaling---allowing the scaled instance 
to increase system throughput before all parameters are fully loaded---is 
necessary because
SLO violations can still happen ({\fig{fig:motiv-scale-speed}} (c)) even with fast networking. 
It is challenging to achieve this in existing systems. 
For example, 
PipeSwitch~\cite{pipeswitch} and DeepPlan~\cite{deepplan} leverage the layer-by-layer 
execution nature of models to perform inference: 
As shown in {\fig{fig:overview}} (c), 
once the first layer is loaded on inst.1, 
they redirect the overloaded requests to it for execution. 
Meanwhile, inst.1 will load subsequent layers concurrently. 
However, such overlapping is not live 
because, until all the layers are loaded, inst.1 still cannot finish requests
to increase the system throughput.

To this end,
we propose a novel cooperative execution scheme for live autoscaling.
The key observation is that though the scaled instance alone
cannot finish the requests until all layers are loaded,
it can alleviate the load of the overloaded instance with its loaded layers,
thus improving the serving throughput.
{\fig{fig:overview}} (d) illustrates this.
When inst.0 becomes overloaded and inst.1 is under scaling,
after inst.1 has loaded the first layer,
we redirect all requests from inst.0 to inst.1 for execution.
Once inst.1 completes the first layer's execution,
it forwards the activation back to inst.0 to process the remaining layers,
and the system throughput increases with reduced queued latencies,
as queued requests are processed faster. 
To see why the throughput increases, 
consider serving a 7-layer model.
inst.0 alone will have a throughput of 1/7.
With our live scaling,
after loading one layer on inst.1,
inst.0 only needs to execute 6 layers,
so its throughput increases to 1/6.
The throughput continues to improve as more layers are loaded,
reaching the peak (doubled) after half of the layers have been loaded---half of the scaling time.
\textsection{\ref{sec:design-pp}} describes our ZigZag scheduling for coordinating overloaded
and new instances during live autoscaling to achieve optimal performance for live autoscaling.

\section{Detailed Design and Implementation}
\label{sec:design}

\subsection{Online network-based scale plan generation}
\label{sec:design-plan}
    
\noindent
When the planner is notified to scale the parameters onto $n$ new GPUs,
it will get $s$ sources from the parameter pool, find $t$ spare GPUs as target
and generate a plan on how to send parameters from $s$ sources to 
a subset of $n$ GPUs in $t$ targets.
There are three metrics to minimize for the generation:
(1) the scale time,
(2) the plan generation time, and
(3) the interference with serving workloads.
\begin{figure}[!t]
        \begin{minipage}{1\linewidth}
        \hspace{-1mm}
        \centering    
        \includegraphics[width=1\textwidth, trim=0.25cm 11.45cm 23.1cm 0.25cm, clip]{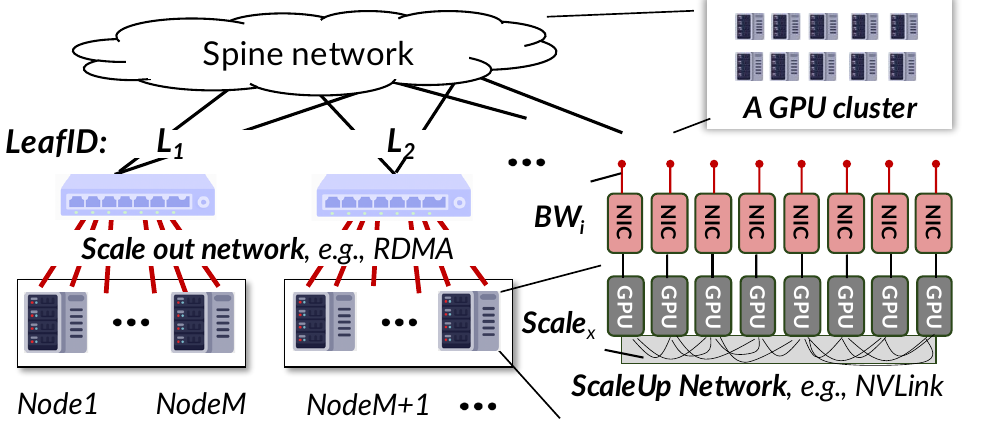} 
        \end{minipage} \\[2pt]
        \begin{minipage}{1\linewidth}
        \caption{\small{\emph{
            An illustration of how {\sys} models the network.
        }}}
        \label{fig:network-topo}
        \end{minipage} \\[-20pt]
        \end{figure} 

\stitle{Modeling the network between GPUs.\,}
Effectively generating a plan requires a model of the network between sources and targets,
which is non-trivial due to the complexity of the network topology in serving clusters.
Our model assumes a simplified scale-up and scale-out network hybrid networking,
widely adopted for GPU clusters~\cite{dellAiFabric, DBLP:conf/hoti/WangGSZH24, nvidia-compute-network}.
{\fig{fig:network-topo}} (a) illustrates the modelling. 

First, we model the GPUs connected via fast scale-up networking like NVLink
as groups of GPUs.
Such GPUs have ultra-high interconnect bandwidth (1,600-3,600\,Gbps)
so scaling within a group has negligible overhead.
On the other hand, GPUs connected via slower scale-out networking like RDMA
are more difficult to model due to a hierarchical structure.
To this end, we adopted a simple leaf-to-spine model that covers
widely deployed topologies including Clos and Rail-optimized~\cite{DBLP:conf/hoti/WangGSZH24} with different subscription ratios:
each GPU ($i$) has a $BW_i$ bandwidth connected to a leaf switch ($\text{LeafID}$),
where GPUs within the same leaf switch have a full-mesh connection,
i.e., the bandwidth between GPUs ($i$) and ($j$) is $\text{min}(\text{BW}_i, \text{BW}_j)$ with full bandwidth.
Second, leaf switches are connected to spine switches,
with inter-leaf bandwidth equal to or smaller than the intra-leaf bandwidth.
For simplicity, we don't model the spine network and rely
on upper-tier protocols like Virtual Link Trunking (VLT~\cite{DellVLTPatent2016}) and Equal-Cost Multi-Path (ECMP~\cite{RFC2992}).
to support efficient inter-leaf networking.

\begin{figure}[!t]
        \begin{minipage}{1\linewidth}
        \hspace{-1mm}
        \centering    
        \includegraphics[width=1\textwidth, trim=0.25cm 14cm 21cm 0.25cm, clip]{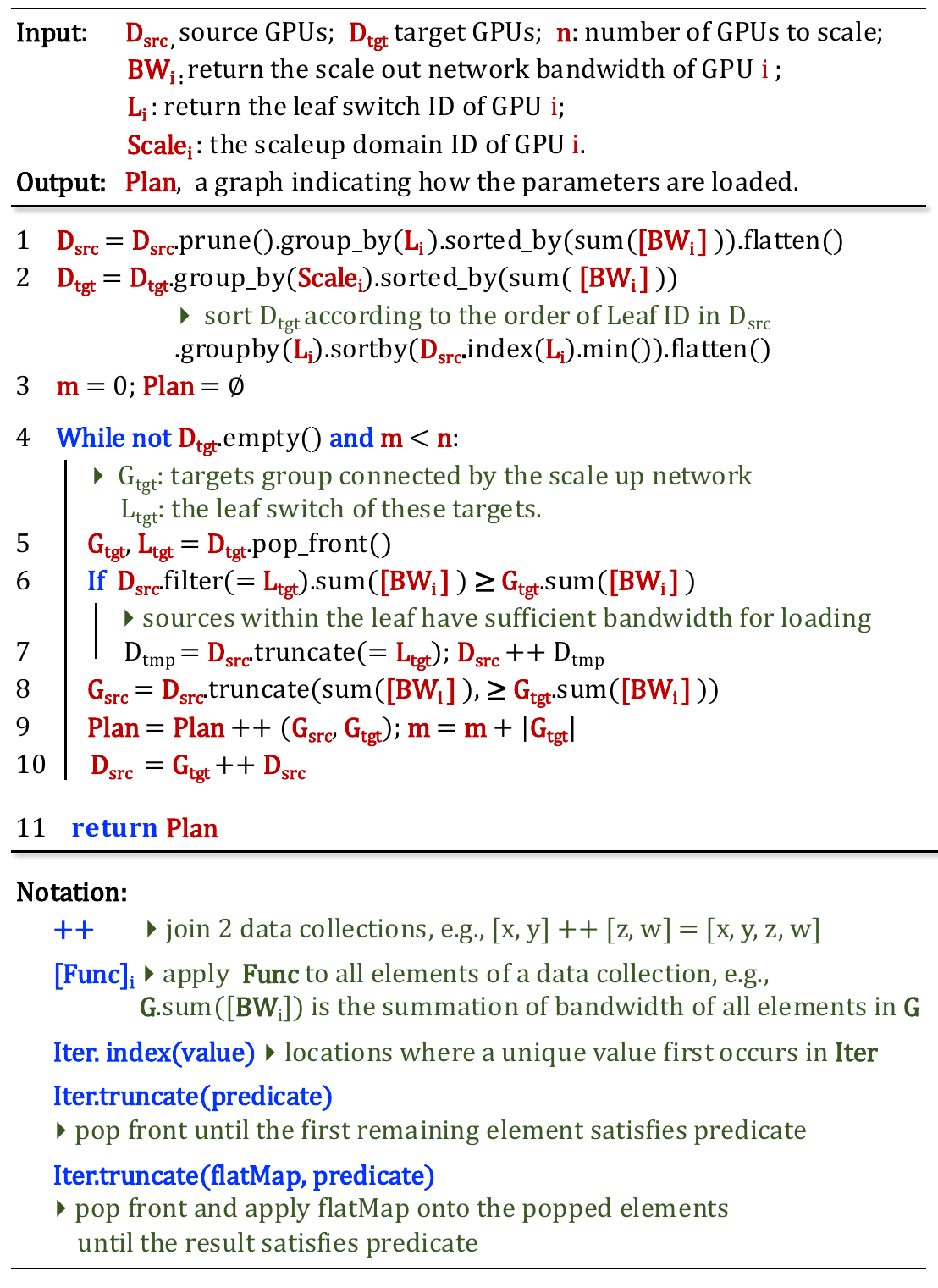} \\[1pt]
        \end{minipage} \\[3pt]
        \begin{minipage}{1\linewidth}
        \caption{\small{\textit{
            The pseudocode of the plan generation algorithm.
        }}}
        \label{alg:alg-plan}
        \end{minipage} \\[-15pt]
        \end{figure}         

\begin{figure}[!t]
        \begin{minipage}{1\linewidth}
        \hspace{-1mm}
        \centering    
        \includegraphics[width=1\textwidth, trim=0.25cm 14.2cm 22.5cm 0.25cm, clip]{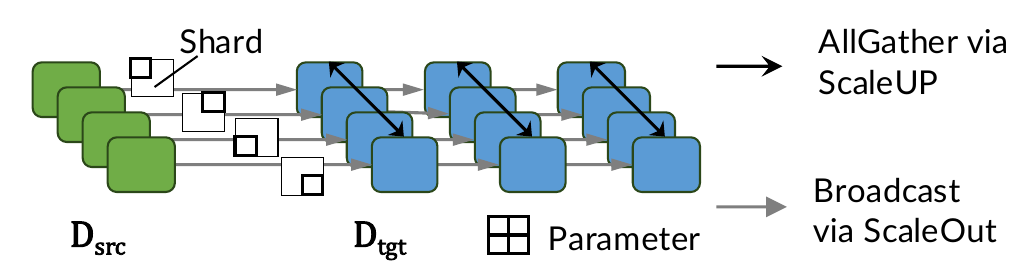} \\[1pt]
        \end{minipage} \\[4pt]
        \begin{minipage}{1\linewidth}
        \caption{\small{\emph{
            An illustration shareded parameter transfer with scale up network. 
        }}}
        \label{fig:parallel-shards}
        \end{minipage} \\[-15pt]
        \end{figure}

\stitle{Multicast-chain-based greedily plan generation. \,}
To quickly generate the plan online, we use a three-step greedy algorithm
as shown in Algorithm~\ref{alg:alg-plan}.
First, we prune the sources to avoid any interference with serving workloads (Line 1).
Second, we group targets connected with scale-up networking like NVLink as
a group (Line 2) such that we can leverage the NVLink broadcast
to efficiently realize parallel sharded parameter transfer,
see {\fig{fig:parallel-shards}} and described below.
Finally, we form multiple serial broadcast chains (Line 3--10) 
to generate the plan.

\begin{figure}[!t]
        \begin{minipage}{1\linewidth}
        \hspace{-1mm}
        \centering    
        \includegraphics[width=1\textwidth, trim=0.25cm 11.6cm 23cm 0.25cm, clip]{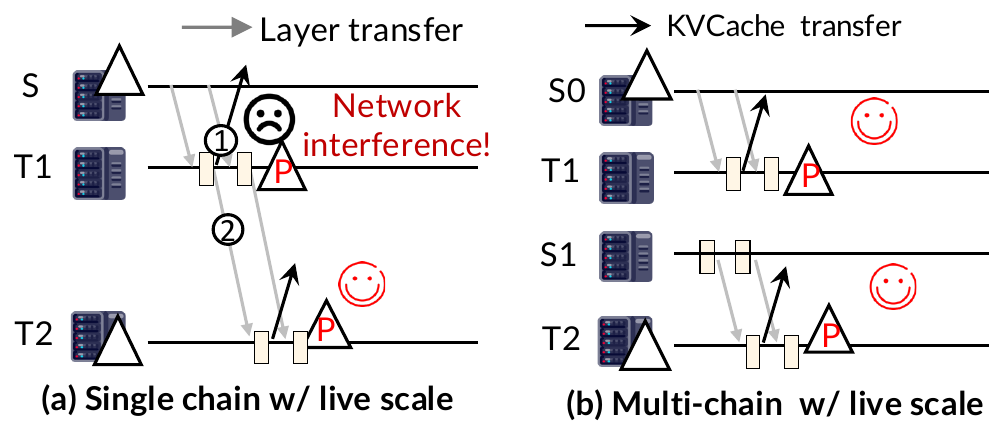} \\[1pt]
        \end{minipage} \\[4pt]
        \begin{minipage}{1\linewidth}
        \caption{\small{\emph{
            An illustration of why multiple chains are better especially 
            under live scaling.
        }}}
        \label{fig:ring-2}
        \end{minipage} \\[-15pt]
        \end{figure}          

\begin{figure}[!t]
        \begin{minipage}{1\linewidth}
        \hspace{-1mm}
        \centering    
        \includegraphics[width=1\textwidth, trim=0.25cm 11.6cm 17.5cm 0.25cm, clip]{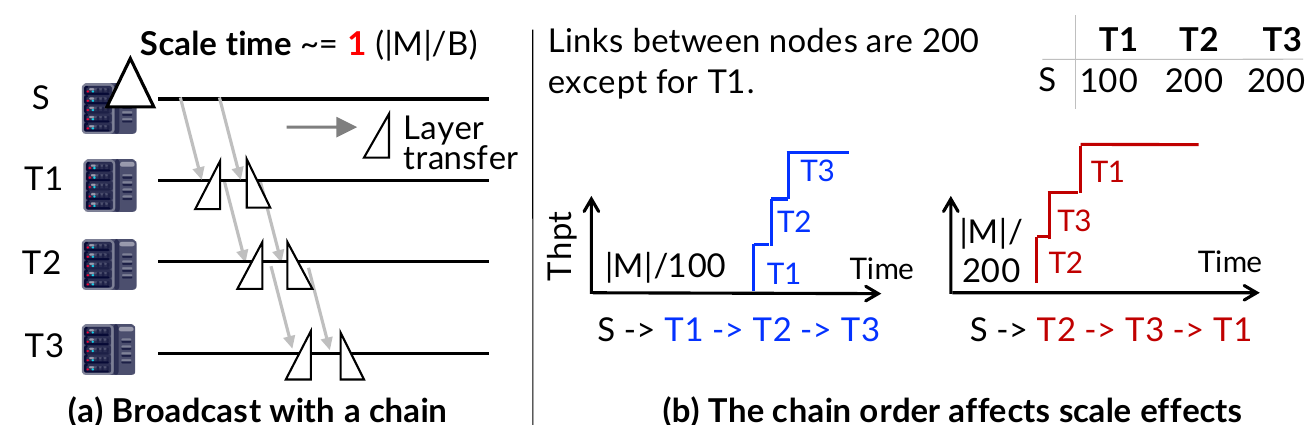} 
        \end{minipage} \\[7pt]
        \begin{minipage}{1\linewidth}
        \caption{\small{\emph{
            An illustration of 
            (a) why chain is friendly to broadcast with huge bandwidth requirement and
            (b) why we chooses a specific chain order. 
            $|M|$ is the model size and $B$ is the slowest network bandwidth between 
            nodes in a chain.
        }}}
        \label{fig:ring}
        \end{minipage} \\[-10pt]
        \end{figure} 

Specifically, a serial broadcast chain is formed by a set of source and target nodes,
i.e., $S \rightarrow T_1 \rightarrow T_2 \rightarrow ... \rightarrow T_n$.       
Note that a node in a chain may have multiple GPUs. 
Such a chain has a nice property that it is optimal in bandwidth-intensive transfer like 
model scaling, because the overall transmission time is irrespective of the instances scaled with such a chain.
As shown in {\fig{fig:ring}} (a), 
when T1 receives the first layer,
T1 immediately forwards it to T2.
Meanwhile, $S$ will continue to send the second layer to T1,
so the time of sending the first and second layer is overlapped.

While a serial chain is sufficient for efficient parameter broadcasting for nodes that are connected
with the same bandwidth links,
multiple chains are necessary in a leaf-spine network where inter-leaf bandwidth may vary
and in our live scale setup.
This is
because (1) multi-chain avoids relatively slow inter-leaf communications if each leaf switch
has sources and targets (Line 6--7),
and (2) it enables more interference-free live scaling especially in PD disaggregation setting.
{\fig{fig:ring-2}} illustrates the latter:
suppose we want to scale up two prefill instances in a live manner,
the KVCache will be transferred to the decode instances once prefill is done.
With a single chain, only $T2$ can join the live scale without network interference,
because at $T1$, the KVCache transfer (\ding{192}) interferes with the
parameter forward traffic (\ding{193}).
With two chains backed by two parameter sources (b), both $T1$ and $T2$
can live scale without interference.

Note that the order of nodes in the chain is important:
we chose a decreasing order with respect to the aggregated link speed between nodes (Line 2, 5).
This is because 
sending to nodes with higher bandwidth achieves a faster increase in the serving throughput. 
As shown in {\fig{fig:ring}} (b): 
suppose the source ($S$) sends parameters to $T2$ twice as fast as $T1$.
A chain order of $S \rightarrow T2 \rightarrow T1$ is better than $S \rightarrow T1 \rightarrow T2$
because the downtime of $T2$ is only half.
Note that the source can be a group of GPUs because GPUs typically have dedicated network cards
in our setup.

\stitle{Optimization: parallel sharded parameter transfer from multiple sources. \,}
\label{sec:design-shard}
For a broadcast chain where the source and target contain GPUs with duplicated parameters,
we further leverage the scale-up network to parallelize a transfer link.
{\fig{fig:parallel-shards}} shows a concrete example.
Suppose the source and target nodes have four GPUs each.
For such a transfer, each source GPU only needs to forward 1/4 of the sharded parameters to each target GPU,
where the target GPUs can use NVLink-based AllGather to get the full parameters.
This reduces the scaling time to 1/4 as the NVLink AllGather time is negligible.

\subsection{Efficient live autoscaling with ZigZag scheduling}
\label{sec:design-pp}

\nospacestitle{Selecting instances for live scaling. \,}
After getting the chains from Algorithm~\ref{alg:alg-plan},
we select instances to participate in live autoscaling
based on the following criteria:
(1) the necessity of live autoscaling,
i.e., when a stop-the-world scaling will cause SLO violation and
(2) the presence of overloaded instance that can cooperate.
Both are readily available:
(1) we can profile the relationship between load speed and SLO violation
 in {\fig{fig:motiv-scale-speed}} for the judgement and
(2) autoscaling is typically triggered when the system is overloaded.
Thus, for each overloaded instance, we
will identify an instance in the chain that satisfies (1),
typically the tail instances in the chains as it has the slowest link.

\stitle{Live autoscale protocol with paired instances. \,}
Suppose we have selected a new instance (inst.1) to offload computations 
from an overloaded instance (inst.0). 
To begin live autoscaling, 
we use a three-step transition protocol:
(1) Once inst.1 starts loading parameters, 
we redirect all queued and new requests from inst.0 to it for execution. 
The redirection time is negligible because the request payloads are much smaller than the model. 
(2) After the first layer is loaded on inst.1, 
it begins executing the first layer of all requests. 
Note that during the loading of the first layer, 
inst.0 remains active by processing its pending requests. 
Finally, when the model has completed loading on inst.1 (3), 
requests will be re-distributed evenly between 
both instances.

\begin{figure}[!t]
        \begin{minipage}{1\linewidth}
        \hspace{-1mm}
        \centering    
        \includegraphics[width=1\textwidth, trim=0.25cm 7.1cm 14.2cm 0.25cm, clip]{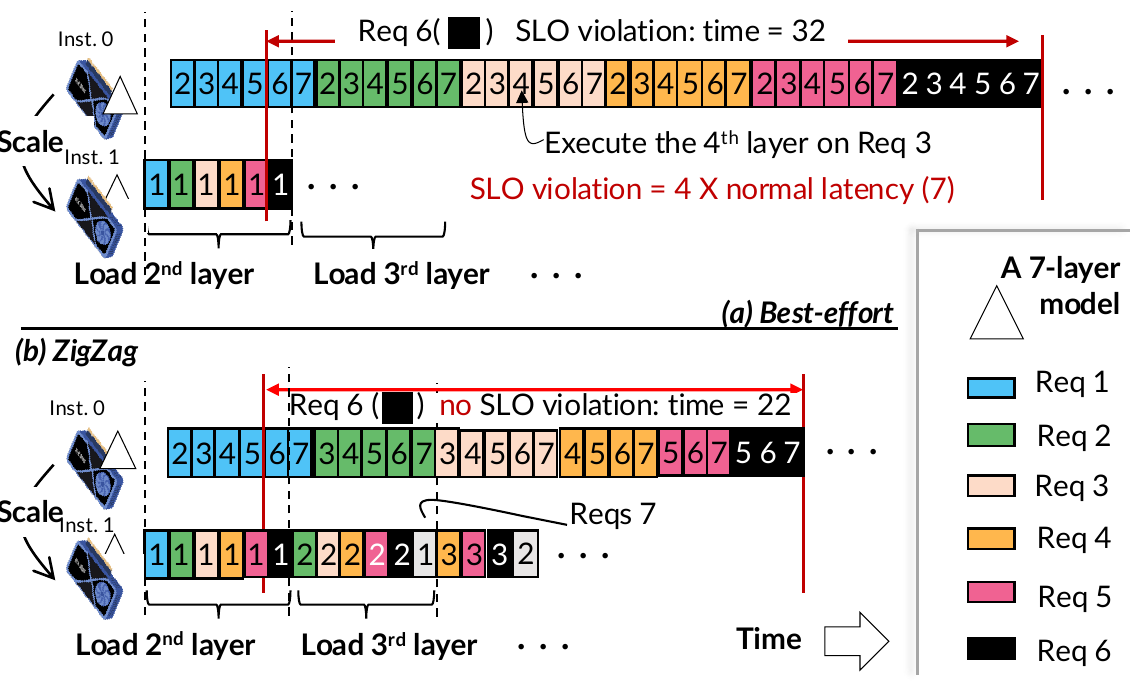} \\[1pt]
        \end{minipage} \\[3pt]
        \begin{minipage}{1\linewidth}
        \caption{\small{\textit{
    An illustration of the necessity of ZigZag scheduling.
    Note that the execution starts when the first layer has been loaded to instance 1 (inst.1).
    Our example assumes the time of loading a layer can perform 6-layer computations.
        }}}
        \label{fig:zigzag}
        \end{minipage} \\[-15pt]
        \end{figure} 

\stitle{The scheduling problem. \,}
A key issue to address in the above (3) is 
how to best utilize inst.1 to maximize the goodput during live autoscaling. 
Specifically, we should decide a pipeline configuration for each request batch,
i.e., how many layers to execute on inst.1 and inst.0, respectively.
One naive policy is best-effort:
for each batch, we execute as many layers as possible on inst.1 (not exceeding half)
and execute the rest on inst.0.
While it adapts configurations with model loading,
we found it is suboptimal because
inst.1 has limited serving capacity during initial loading,
so most requests are still queued at inst.0, causing SLO violations.
{\fig{fig:zigzag}} (a) shows a concrete example
of a 7-layer model executed with the best-effort scheduling.
The load time of one layer can do 6-layer computations,
a common setup (e.g., Llama2-7B model with a moderate batch size of 2000 prefill tokens under 200\,Gbps RDMA network).
Thus,
before the second layer has been loaded to inst.1,
the current request batches (req 1--6) can only use a $(1,6)$ pipeline configuration.
However, request 6 will suffer from SLO violation
due to waiting for requests 1--5 to be completed on inst.0,
as their execution time has reduced a little due to the imbalanced loads.

\stitle{The ZigZag scheduling. \,}        
To address this issue, our observation is that
by delaying request scheduling on inst.0,
inst.1 will have more layers coming,
opening opportunities to balance the workload between instances.
{\fig{fig:zigzag}} (b) shows this:
After requests 2--5 have been executed on inst.1,
we delay their execution on inst.0 and wait for the second layer to come.
This allows us to adopt a more aggressive pipeline configuration $(2,5)$ for them.
Note that the delay won't waste GPU because we can schedule pending requests (e.g., 6).
After the second layer has been loaded,
inst.1 can come back (thus, in a ZigZag way) to execute the second layer of requests 2--5.
Thus, the second layer execution of requests 3--6
is overlapped with the execution of layers 3--6 for request 2.
Thanks to this overlap, the overall inference time of request 7
decreases from 32 to 22, now within the SLO.

The above ZigZag scheduling can be formulated as follows.
Assuming a first-come-first-serve (FCFS) scheduling policy,
the de facto for serving~\cite{vllm,DBLP:journals/corr/abs-2404-09526,DBLP:conf/osdi/YuJKKC22}.
For ease of presentation,
we first assume non-LLM and then extend to LLM in \textsection{\ref{sec:design-llm}}.
The scheduling has two parts:

\emph{\uline{(1) Pipeline configuration. }}
Given $N$ request batches with equal execution time,
we first determine the pipeline configuration ($\text{T}_i, \text{S}_i$) for them,
where $\text{T}_i$ and $\text{S}_i$ are the number of layers
to be executed on the target and source GPU for request $i$,
respectively.
The goal is to minimize the average latency,
which can be formulated with the following Integer Linear Program (ILP): \\[-5pt]
\[
 \text{Latency}_\text{avg} = (\sum_{req=1}^{N} \sum_{i=1}^{req} \text{S}_{i}) / N
\] \\[-3pt]
\noindent
To see why such a formula holds, consider the example in {\fig{fig:zigzag}} (b).
Each request's latency is the time the source instance finishes its part of the computation,
which includes its own execution time and the sum of its previous requests' time (queueing time).
We only need to consider previous requests because they are executed in a FIFO order.
In our example, request 3's latency is 17 (12 for requests 1 and 2's execution and
5 for its own).
For non-LLM, the execution time of each layer is the same if the batch size is the same.
Note that we omit the activation transfer latency since it is negligible.

The problem has the following constraints: 
\begin{mini}
    {}{\text{Latency}_\text{avg}}{}{} \\[-20pt]
    \addConstraint{\text{S}_{i} +  \text{T}_{i} = L, \quad \forall i \quad \text{Pipeline limit \bf{(C1)}}}{}  \vspace{-10pt}
    \addConstraint{\sum_{j=1}^{i} \text{T}_j  \le \sum_{j=1}^{i-1} \text{S}_j,\forall i > 1 \quad \text{Pipeline dependency \bf{(C2)}}}{} \vspace{-10pt}
    \addConstraint{\text{Time}_{l} \ast \text{T}_i \leq \sum_{j=1}^{i-1} \text{T}_j + (N-i+1) \times (\text{T}_i - 1)} \vspace{-15pt}
    \addConstraint{\forall i > 1 \quad \text{Load limit \bfseries (C3)}}{}
    \label{eq:pp-config}
  \end{mini}
\noindent
\textbf{C1} ensures that the pipeline should be fully executed. 
\textbf{C2} states pipeline dependency: 
when the source instance executes request $i$'s $S_i$ layers, 
the target instance must finish the execution of $T_i$. 
The start execution time of request $i$ on the source is 
$\sum_{j}^{i-1} \text{S}_j$.
The finish time of $i$ on the target instance is $\sum_{j}^{i-1} \text{T}_j + \text{T}_i$,
which simplifies to $\sum_{j}^{i} \text{T}_j$. 
Finally, \textbf{C3} ensures that once the target instance request's $i$'s $T_i$ layers, 
all these layers must be loaded, where $\text{Time}_l$ is 
the time to load one layer normalized to the execution time of one layer in pipeline. 
The term $(N-i+1) \times (\text{T}_i - 1)$ indicates that the load time can be 
overlapped with executing of the succeeding requests of $i$. 

While solving this ILP is NP-hard, it remains manageable (less than 40\,ms to solve for Llama3-8B)
because models typically have only a few dozen layers. 
Additionally, we only need to configure the pipeline for the batches of requests 
executed during parameter loading, which is a dozen of so in practice.
Nevertheless, 
to further eliminate the solving time for models with more layers (e.g., 80 layers for Qwen-72B),
we also derived an ILP-free method that we described below. 

\begin{figure}[!t]
        \begin{minipage}{1\linewidth}
        \hspace{-1mm}
        \centering    
        \includegraphics[width=1\textwidth, trim=0.25cm 6.99cm 21.3cm 0.25cm, clip]{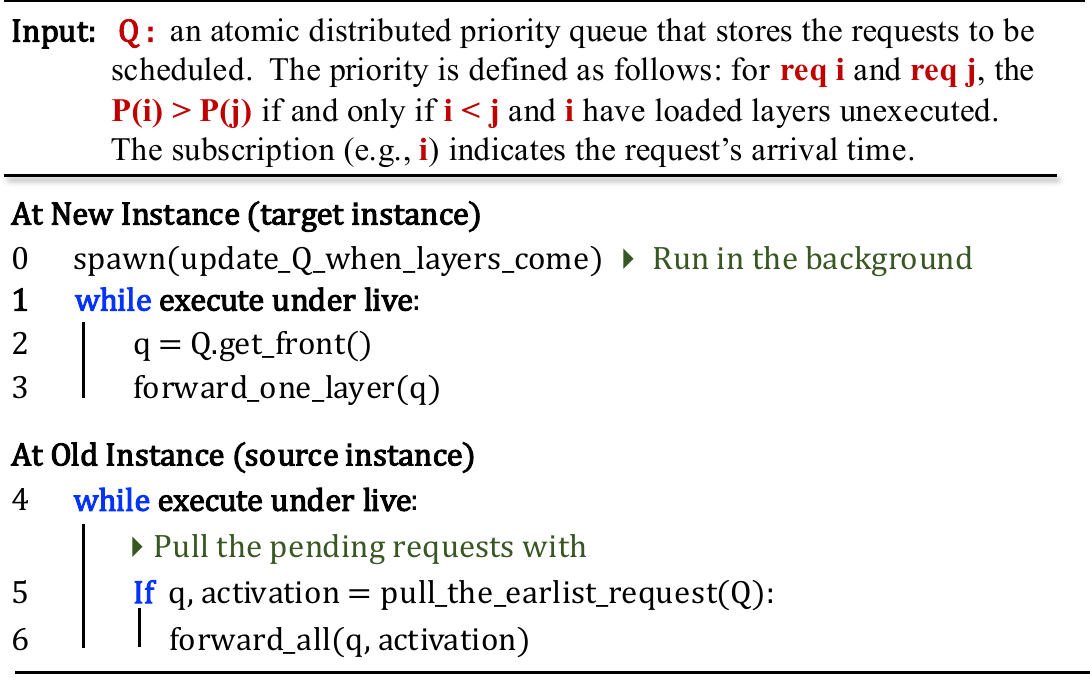} \\[1pt]
        \end{minipage} \\[3pt]
        \begin{minipage}{1\linewidth}
        \caption{\small{\textit{
            The pseudocode of the ILP-free ZigZag scheduling.            
        }}}
        \label{fig:zigzag-algo}
        \end{minipage} \\[-15pt]
        \end{figure} 

\emph{\uline{(2) Scheduling requests in an ILP-free ZigZag way. }}
Specifically, we found that by delaying sending the requests on the
source instance
and letting the target instance execute the requests once it is free,
we can achieve ZigZag scheduling without solving the ILP.
{\fig{fig:zigzag-algo}} shows the pseudocode of how we schedule the requests on both instances.
The new instance maintains a priority queue (that can be pulled by the source instance via RPC)
for all requests,
where the priority is defined by (1) the FCFS order and
(2) requests with next-to-execute layer loaded coming first.
Once the new instance has executed one layer (Line 3),
we keep the executed request in the queue so it can be scheduled back once
more layers are loaded.
The requests are scheduled on the source instance only if
it is not overloaded, i.e., has no pending requests (Line 5).
Thus, if the source instance is busy,
the request will still be executed on the target instance.

\subsection{Global parameter pool and scaling policy}
\label{sec:design-others}

\nospacestitle{Global parameter pool and local memory cache. }
Our global parameter pool tracks the locations of the model parameters across deployed GPUs and host CPUs
with local memory cache.
To ensure at least one copy of the model parameters is available in the
memory of GPU or host at the cluster scale,
during system initialization,
we distribute one copy of the model's parameters evenly
to the CPU hosts and track their locations at a centralized manager.
When a model is deployed to or reclaimed from a GPU,
we further update the locations in the manager,
and reclaim/reload cached copies on the host cache.

\stitle{Scaling policy. \,}
Our paper focuses on the autoscaling mechanism,
which is orthogonal to the autoscaling policies,
including
collecting workload metrics with workload monitoring
and determining how many new instances to scale based on these metrics.
Our current implementation follows prior works~\cite{DBLP:conf/osdi/SunHZXZL024,K8S-HPA}
that first records the serving loads with tokens per second and KVCache usage globally.

For scaling up,
when the average monitored load surpasses a pre-defined upper bound,
we allocate sufficient instances to meet that demand.
The upper bound can be derived by profiling the average serving load per-instance offline.
We leave a more detailed explanation in another paper. 
For scaling down,
we follow previous works'\cite{DBLP:conf/usenix/Romero0YK21,DBLP:journals/corr/abs-2401-14351}
timeout-based policy:
when the average monitored load falls below a lower bound in a time window,
we shut down some instances and revoke all GPUs assigned to them.
Given {\sys}'s rapid autoscaling capabilities,
we adopt an extremely short sub-second level timeout.

\subsection{Specializations and optimizations for LLM}
\label{sec:design-llm}

\noindent
While most techniques described above work for all models following a
layer-by-layer architecture,
the unique characteristics of LLMs especially for LLMs served with PD disaggregation
require several specializations and optimizations.

\stitle{Retrofitted live pipeline scheduling formulation. \,}
Our formulas and constraints described in \textsection{\ref{sec:design-pp}} cannot be directly applied
to LLMs
because the prefill and decode time of a layer is approximately
linear to the total batched token size~\cite{DBLP:conf/isca/PatelCZSGMB24,DBLP:journals/corr/abs-2404-09526}.
For prefill-only live scheduling, e.g., autoscaling a prefill instance in PD disaggregation,
we fix the formulation by adding
a regulation parameter for each request batch by profiling its execution time with the counts,
similar to a priori work~\cite{DBLP:journals/corr/abs-2404-09526}.
A more tricky case involves handling decoding,
e.g., when scaling instances that combine prefill and decode,
or scaling a decode instance in PD disaggregation.
The complexity arises because decode batch size changes dynamically
due to its auto-regressive nature.
Fortunately, our ILP-free scheduling method can also work for decoding.

\stitle{Supporting PD colocation. \,}
We seamlessly support PD colocation since a PD-colocated instance 
is a normal model instance. 
Meanwhile, our ILP-free ZigZag scheduling also applies to pipelined execution under PD colocation~\cite{298679}.

\stitle{Live scaling decode instances in PD disaggregation. }
Live scaling a decode instance in PD disaggregation without interference is impossible
due to the incast bandwidth contention of both parameter loading and KVCache transfer.
Thus, we leverage the fact that the prefill and decode instances share the same model parameters,
so we can live scale a decode instance by first mutating some
prefill instances to decode instances,
while concurrently live scaling the prefill instances to compensate for the prefill throughput.

\stitle{Optimized scaling policy for PD disaggregation. \,}
Pre-scaling instances can hide the cost of scaling,
but a too-early scaling wastes GPU resources.
In PD disaggregation,
we found we can pre-scale decode instance at zero cost,
because the need for scaling decode instances
can be evidenced by the requirement for scaling prefill instances.
Specifically, once we found a significant requirement for scaling prefill instances,
we will simultaneously scale decode instances.
This effectively hides the scaling cost of decode instances,
and is even effective for other systems like ServerlessLLM~\cite{DBLP:journals/corr/abs-2401-14351},
see \textsection{\ref{sec:eval-app}}.

\section{Evaluation}
\label{sec:eval}

\nospacestitle{System implementation. }
{\sys} is a {\maas} system capable of serving both traditional models and LLMs
with 24,000 lines of Rust and C++ code.
It builds upon widely applied LLM optimizations like PD disaggregation and continuous batching.
We leverage existing highly-optimized serving system components (with no autoscaling support) wherever possible.
For instance,
all our GPU kernels for LLM come from FlashInfer~\cite{flashinfer}.
We choose a native-language-based framework implementations because we found 
it is challenging to implement fine-grained scheduling in Python.
\textsection{\ref{sec:appendix}} provides more implementation details.

\begin{table}[!t]
    \centering
    \small{
        \resizebox{.97\linewidth}{!}{
            \ra{1.2}

            \begin{tabular}{lrr} \toprule
                                 & \textbf{Cluster A} ($m$ x $g$)     & \textbf{Cluster B} ($m$ x $g$)     \\ \hline
                GPU              & A800 80\,GB (4x8)           & A100 80\,GB (2x8)         \\
                GPU-GPU (intra)  & 1.6\,Tbps NVLink & 256\,Gbps PCIe \\
                GPU-GPU (inter)  & 100\,Gbps RDMA  & 100\,Gbps RDMA \\
                Host-GPU         & 128\,Gbps PCIe  & 128\,Gbps PCIe \\
                SSD-GPU          & 10\,Gbps        & 10\,Gbps       \\ \bottomrule
                \end{tabular}        
            
        }
    } \\[8pt]
    \begin{minipage}{1\linewidth}
        \caption{\small{\textit{
        Evaluation clusters. $m$ is the number of hosts and 
        $g$ is the number of GPUs per host.
        }}}
    \label{tab:cluster}
    \end{minipage} \\[-10pt]
\end{table}

\stitle{Testbed. }
Our evaluations are conducted on two testbeds listed in Table~\ref{tab:cluster}.
Cluster A can serve larger models (e.g., 72\,B) with tensor parallelism~\cite{vllm}
thanks to the NVLink 
while cluster B is more suitable for serving single-GPU models. 

\stitle{Evaluated traces and models. }
Because the scaling requirements are closely related to the incoming request rates,
we chose three typical real-world traces:
BurstGPT~\cite{wang2024burstgpt},
AzureCode and AzureConv, both from Azure~\cite{DBLP:conf/isca/PatelCZSGMB24}.
The detailed trace shapes are shown in the first column in {\fig{fig:eval-e2e}}.
Since the traces are collected from clusters with different serving capabilities,
we follow the standard approach~\cite{DBLP:conf/asplos/MiaoSDXL0J24,DBLP:journals/pvldb/AliPYS22,DBLP:conf/osdi/GujaratiKAHKVM20}
to scale the traces to fit our clusters.
Specifically, we scale the trace with temporal pattern preserved using
TraceUpscaler~\cite{traceupscaler},
and the scaled 
average request rate is half of the maximum serving capacity of our cluster.

For models, we focus on evaluating LLMs because other non-LLMs
are much smaller and trivially scale efficiently with {\sys}.
Specifically,
we choose Llama3-8B, Mistral-24B and Qwen2.5-72B,
all are popular LLM models with high accuracy.
Since {\sys} is only sensitive to the model size,
we may omit the detailed model family name and only uses their sizes
in the following description for simplicity.
For small model (8B), it only needs one GPU per instance while
for 72\,B models the minimal number of GPUs used by one instance is 4.

\stitle{Comparing targets. \,} 
Without explicit mention,
we compare {\sys} with the following baselines:
\begin{enumerate}[leftmargin=*]
    \itemsep0.5em    
    \setlength{\itemindent}{0em} 
    \item \textbf{\emph{ServerlessLLM (S-LLM)
    ~\cite{DBLP:journals/corr/abs-2401-14351}}} 
    is the state-of-the-art {\maas} with a focus on accelerating autoscaling speed.
    It utilizes host memory to cache recently loaded models with a time-to-live eviction policy.
    Under cache misses, it loads parameters from SSD with SSD bandwidth fully utilized. 
    \\[-15pt]

    \item \textbf{\emph{ServerlessLLM optimal (AllCache)}}
    is the autoscaling speed optimal version of ServerlessLLM that always 
    loads the parameters from the host cache. \\[-15pt]

    \item \textbf{\emph{DistServe}}~\cite{DBLP:conf/osdi/ZhongLCHZL0024}
    is the state-of-the-art LLM serving system without autoscaling support.
    It leverages PD disaggregation.
    We chose it because autoscaling is more challenging in
    PD disaggregation due to the complexity of multiple instances scaling (prefill and decode)
    and the need to avoid scaling interference.
    We compare with other common PD colocation systems like vLLM~\cite{vllm} in \textsection{\ref{sec:eval-pd-colocation}}. \\[-15pt]
\end{enumerate}    

\noindent
For a fair comparison, 
we adopted the same scaling policy for both {\sys} and variants of S-LLM.

\subsection{Autoscaling performance under real-world traces}
\label{sec:eval-app}

\begin{figure*}[!t]
    \centering
    \includegraphics[width=1.01\textwidth,center]{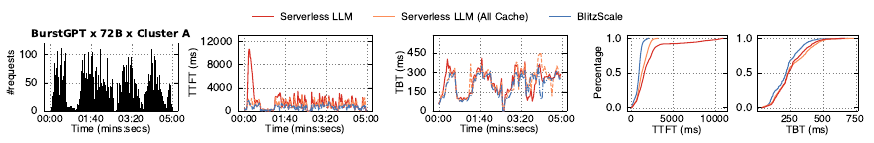} \\[0pt]
    \includegraphics[width=1.01\textwidth,center]{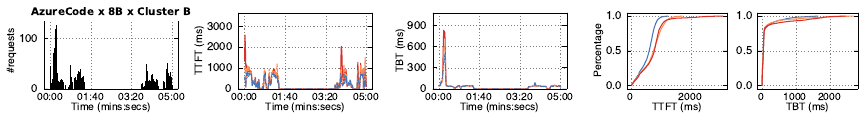} \\[0pt]
    \includegraphics[width=1.01\textwidth,center]{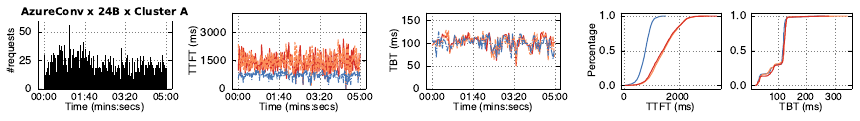} \\[-5pt]
    \begin{minipage}{1\linewidth}
    \caption{\small{\textit{
        End-to-end performance comparison between {\sys} and {\base} on various workloads,
        models and clusters. 
    }}}
    \label{fig:eval-e2e}
    \end{minipage} \\[-10pt]
\end{figure*}

\noindent
Due to space limitations,
for each model, we choose one trace on one of the clusters to evaluate the performance.
{\fig{fig:eval-e2e}} presents the end-to-end performance
when serving with a prefill and decode disaggregation setup
where the instances for different phases are scaled independently.
The first column shows the request rate of the trace,
the second and third columns show the mean TTFT and TBT, respectively,
where each point is the average latency measured during a small time window (1s),
and the final two columns present the cumulative distribution function (CDF) of the TTFT and TBT
during the evaluation period, respectively.
We focus on comparing with S-LLM and AllCache in this section
and leave the comparison with DistServe in the next section,
as it does not support autoscaling.

\stitle{Overall performance. }
First, we can see that {\sys} has the lowest TTFT and TBT in all workloads
thanks to the fast autoscaling speed.
Specifically,
on BurstGPT, the TTFT is 75.5\,\% and 21.1\,\% shorter than S-LLM and AllCache,
respectively,
and the TBT is 7.4\,\% and 5.1\,\% shorter, respectively.
Nevertheless,
the degrees of improvement are different across metrics
due to the unique characteristics of the prefill and decode phases.
Meanwhile,
the behaviors of systems, especially S-LLM, are different across workloads
due to the different request arrival patterns (see the first column).
We elaborate on the differences in the following.

\stitle{TTFT vs. TBT. }
{\sys} is more effective in reducing the TTFT than TBT on all workloads.
This is due to two reasons.
First, the decode instance can be pre-scaled 
thanks to our optimized policy (\textsection{\ref{sec:design-llm}}),
which we apply to all baselines. 
Specifically, 
when the prefill throughput increases,
{\sys} (and its baselines) will simultaneously scale the decode instances,
yet no more decode instances are needed at the scale time.
Thus, the scaling time is overlapped with the prefill time, 
which hides some scaling overheads.
Second,
decode scales less than prefill because
as long as there is sufficient memory on the decode instances,
all systems can handle decoding with a slightly increased TBT due to no queueing.
Since all models adopt modern LLM optimization group query attention~\cite{gqa}
with low memory footprint,
decoding instances are more sufficient than prefill instances.
Nevertheless, {\sys} still achieves a 5.1--7.4\,\%, 88.3--94.1\,\% and 0.7--1.8\,\%
shorter TBT than S-LLM and AllCache on three workloads, respectively.

\stitle{Comparisons between different workloads results. }
{\sys} always outperforms AllCache thanks to fast network-based autoscaling
as well as live scaling,
but S-LLM has different behaviors compared to AllCache in these workloads.
On BurstGPT, S-LLM first has a sharp TTFT spike at the first burst (time 0:05),
while it is close to AllCache in future bursts,
because future bursts can benefit from host cache.
In comparison, on AzureCode,
S-LLM has spikes under both bursts (time 0:05 and time 03:25),
because the gap between two bursts makes the host cache evicted due to a time-to-live policy.
Finally, on AzureConv,
since the bursts continuously arrive, S-LLM always hits the host cache,
so the performance---see the CDF graphs---is similar to AllCache.

\subsection{Performance and resource usage}
\label{sec:eval-resource}

\begin{figure*}[!t]
    \centering
    \includegraphics[width=1.01\textwidth,center]{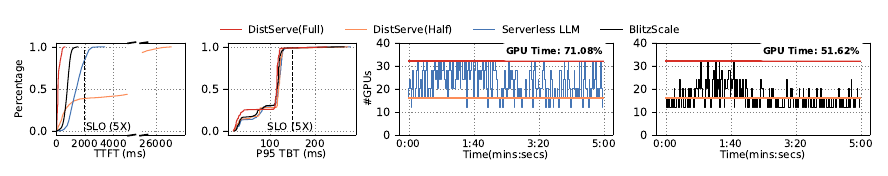}  \\[-10pt]
    \begin{minipage}{1\linewidth}
    \caption{\small{\textit{        
        A comparison between GPU usage under AzureConv with Mistral\,24B.
    }}}
    \label{fig:eval-e2e-gpu}
    \end{minipage} \\[-15pt]
\end{figure*} 

\nospacestitle{Comparison with non-autoscaling systems. }
We first compare {\sys} with DistServe. 
Since DistServe does not support autoscaling,
its performance is highly dependent on the number of provisioned instances.
Therefore, we evaluate two setups:
DistServe (full) uses all GPUs in our cluster and represents
an optimal performance at the cost of GPU waste.
On the other hand, DistServe (half) uses GPUs with the average number of instances
required to handle all the workloads within the evaluation period.
For simplicity, we only present the results on AzureConv on 24B models,
the overall trends are similar.
We have carefully calibrated DistServe's performance,
such that when autoscaling is disabled in {\sys},
DistServe has the same performance as {\sys} in all setups.

The first two columns of {\fig{fig:eval-e2e-gpu}} present the latency results. 
First, it can be observed that DistServe (full) has the best performance,
because the GPU is over-provisioned 
so it doesn't suffer from queueing or scaling overhead.
Nevertheless, {\sys} still achieves the same service level objective (SLO) as DistServe (full)
while S-LLM incurs 18.7\,\% SLO violations.
We follow the traditional 5\,$\times$ SLO~\cite{DBLP:conf/osdi/ZhongLCHZL0024}
since all our workloads (chat and code generation) are latency-sensitive.
Specifically, if a request's end-to-end (TTFT or TBT) latency is exceeds 5\,$\times$ the average latency,
it violates the SLO.
Finally, DistServe (half) has the poorest performance:
on average, {\sys} has a 95.8\,\% and 1\,\% shorter TTFT and TBT than DistServe (half).
{\sys} achieves this by using the same GPU time for serving this model as DistServe (half),
and this time is 50\,\% smaller than DistServe (full),
which we elaborate next.

\stitle{GPU time used. \,}
The last two columns of {\fig{fig:eval-e2e-gpu}} show the GPU time used by S-LLM
and {\sys}, respectively.
For S-LLM and {\sys},
we collected the aggregated GPU usage at each time point for both prefill and decode instances,
and the overall time is calculated by integrating the area under the curve.
For variants of DistServe, their GPU time is constant across the evaluation period.
We can see that {\sys} has 19.46\,\,\% lower GPU time than S-LLM
thanks to the fast autoscaling capability:
with low scaling speed,
there would be more queued requests, so the system would trigger more scaling operations
that use more GPU time.
This is unnecessary with {\sys}.
Even with less GPU time used,
{\sys} has a 48.1\,\% and 1.8\,\% shorter TTFT and TBT than S-LLM, respectively.

\begin{figure}[!t]
    \centering
    \includegraphics[width=1.0\linewidth,center]{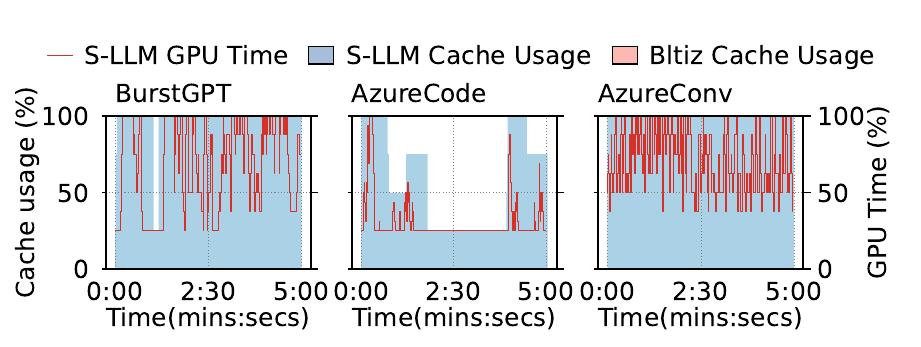}  \\[2pt]
    \begin{minipage}{1\linewidth}
    \caption{\small{\textit{        
        A comparison of host cache usage on S-LLM and {\sys} under the evaluated workloads.
    }}}
    \label{fig:eval-cache-usage}
    \end{minipage} \\[-20pt]
\end{figure} 

\stitle{Host cache usage. \,}
Compared to ServerlessLLM, {\sys}
also consumes less host memory for parameter caching. 
{\fig{fig:eval-cache-usage}} reports the host memory usage for different systems.
We omit AllCache and DistServe,
as AllCache always fully replicates parameters to all hosts while
DistServe does not need caching.
We normalize the host cache usage as different workloads use different clusters.
The results deliver two messages.
First, {\sys} only needs minimal host caching (less than one) to achieve fast autoscaling:
this is as expected by our design because
we prefer to load parameters from GPUs of instances that serve the model,
and even when no serving instance is available,
we only need one host copy due to the network-based multicast.
Second, the memory usage of ServerlessLLM is proportional to the number of hosts involved in the serving,
so a model can quickly ``pollute'' the host cache.
This is non-optimal for an {\maas} system because it can simultaneously
serve many models while the host cache is limited.

\subsection{Detailed performance analysis}
\label{sec:eval-ablation}

\begin{figure*}[!t]
    \centering
    \includegraphics[width=1.01\textwidth,center]{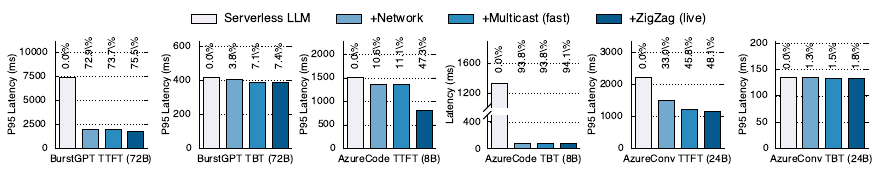}  \\[-2pt]
    \begin{minipage}{1\linewidth}
    \caption{\small{\textit{        
        An ablation study on the effectiveness of our proposed techniques. 
    }}}
    \label{fig:eval-alb}
    \end{minipage} \\[-20pt]
\end{figure*} 

\begin{figure}[!t]
    \centering
    \includegraphics[width=1.0\linewidth,center]{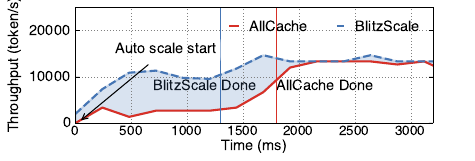}  \\[2pt]
    \begin{minipage}{1\linewidth}
    \caption{\small{\textit{        
    A detailed look at how {\sys} and AllCache scale a 24\,B model to 
    6 prefill instances on cluster A.
    }}}
    \label{fig:scale-breakdown}
    \end{minipage} \\[-5pt]
\end{figure} 

\nospacestitle{A detailed look at the live scale. }
{\fig{fig:scale-breakdown}} shows a throughput timeline
when using {\sys} and AllCache to scale six 24\,B prefill instances.
{\sys} utilizes two broadcast chains (each involving 3 instances),
while the end instances involve a live autoscale. 
The start instances are the decode instances.
For AllCache, it directly loads the parameters from the host memory of the scaled instances.
We can see that first, even with a few loaded layers (e.g., at time 500\,ms),
{\sys} can gradually emit tokens as a result of live execution.
Second, {\sys} can scale faster even compared with AllCache,
thanks to our NVLink-based fused link transmission protocol:
it can finish scaling in 1,200\,ms while AllCache takes about 2,000\,ms.

\stitle{Ablation study. }
{\fig{fig:eval-alb}} conducts an ablation study on the effectiveness of our proposed techniques.
We measured the effectiveness by incrementally enabling different techniques and reporting
the results on the three workloads:
``+Network'' leverages fast computing network instead of SSD for autoscaling,
``+Multicast (fast)'' further applies our optimized parameter broadcast protocol described in \textsection{\ref{sec:design-plan}},
while ``+ZigZag (live)'' enables live autoscaling of \textsection{\ref{sec:design-pp}}.

First, we can see that all techniques are effective in improving the end-to-end serving performance,
but the degrees differ across workloads.
First, ``+Network'' improves the scaling performance in all workloads thanks to the
higher bandwidth for the autoscaling data plane.
Second, ``+Multicast (fast)'' is effective in AzureCode and AzureConv,
but it is less effective in BurstGPT
due to the limitations of our cluster (up to 8 instances can be scaled
on 72\,B model),
so there are no cases to simultaneously scale multiple instances,
which is the targeted case for this technique.
Live autoscaling is mostly effective in AzureCode because it is evaluated on a cluster with
slow networking (Cluster B).
Finally, our techniques are not such effective on decoding because decode instances are sufficient in most cases,
which we have discussed in \textsection{\ref{sec:eval-app}}.
One exception is AzureCode:
in this workload, 
the prefill throughput increases slower than others (see the first column of {\fig{fig:eval-e2e}}),
so the decode instances are triggered later.
As a result, 
the slow scale of ServerlessLLM's SSD cannot be hidden,
making a faster scaling more beneficial.

\begin{figure}[!t]
    \centering
    \includegraphics[width=1.0\linewidth,center]{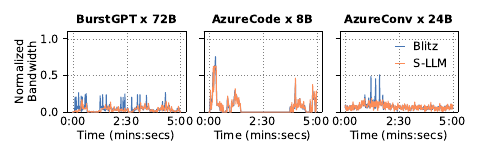}  \\[-10pt]
    \begin{minipage}{1\linewidth}
    \caption{\small{\textit{        
        A profile of the network usage of {\sys} (Blitz). 
    }}}
    \label{ffig:eval-network-util}
    \end{minipage} \\[-10pt]
\end{figure}

\stitle{Network usage. }
{\fig{ffig:eval-network-util}} shows the network usage of {\sys} and S-LLM:
we can see that though {\sys} leverages compute network for autoscaling 
and the scale frequency is high (see the last column of {\fig{fig:eval-e2e-gpu}}),
the additional network usage is negligible.

\begin{figure}[!t]
    \centering
    \includegraphics[width=.95\columnwidth, trim=0.25cm 10.9cm 18.55cm 0.9cm, clip]{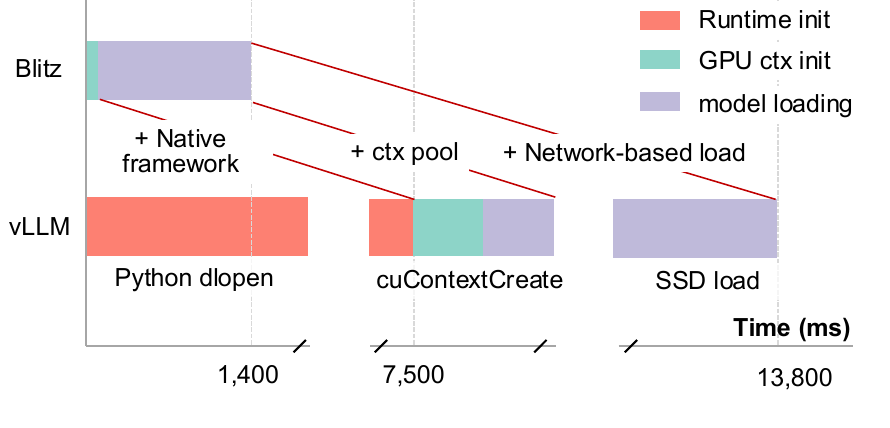}  
    \\[-8pt]
    \begin{minipage}{1\linewidth}
    \caption{\small{\textit{        
        A comparison of init time of {\sys} and vLLM
    }}}
    \label{fig:breakdown}
    \end{minipage} \\[-10pt]
\end{figure}

\stitle{Control plane vs. data plane of model autoscaling. }
{\fig{fig:breakdown}} compares the control plane and data plane overhead during model autoscaling
with vLLM. 
We can see that with proper optimizations,
the control plane overhead is negligible.

\subsection{Performance under LLM PD colocation}
\label{sec:eval-pd-colocation}

\begin{figure}[!t]
    \centering
    \includegraphics[width=1.0\linewidth,center]{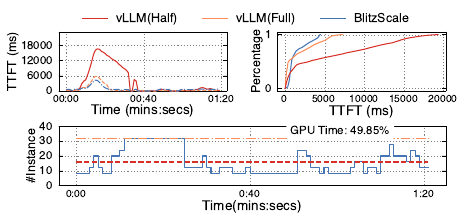}  \\[2pt]
    \begin{minipage}{1\linewidth}
    \caption{\small{\textit{        
        A comparison of {\sys} and vLLM on the BurstGPT workload. 
    }}}
    \label{fig:eval-colocation}
    \end{minipage} \\[-10pt]
\end{figure}


\noindent
Finally, {\fig{fig:eval-colocation}} compares the performance of {\sys} and vLLM
where the serving is conducted in PD colocation on BurstGPT workloads with Llama2-7B model.
The general trend is similar to PD disaggregation:
{\sys} has comparable performance with over-provisioned vLLM,
while compared with an average provisioning,
{\sys} has a 0.24\,$\times$ shorter P99 TTFT.
Interestingly, we found {\sys} has even shorter tail TTFT compared with over-provisioned vLLM,
because our scheduling framework is optimized for cluster serving.

\section{Related work}
\label{sec:related}

\nospacestitle{Optimizing model serving without autoscaling. }
Serving models at scale is non-trivial.
A significant body of research focuses on how to efficiently utilize GPUs to accelerate model serving~\cite{DBLP:conf/osdi/ZhongLCHZL0024,DBLP:journals/corr/abs-2404-09526,vllm,alpaserve,DBLP:conf/osdi/YuJKKC22,DBLP:conf/isca/PatelCZSGMB24,298501,kvcache},
e.g., 
Orca~\cite{DBLP:conf/osdi/YuJKKC22} proposes iterative-scheduling and selective batching. 
AlphaServe~\cite{alpaserve} employs pipeline parallelism to better handle load spikes,
but it cannot adjust pipeline instances dynamically.
These systems assume running on a fixed pool of GPUs,
and we have shown the
necessity of dynamically adjusting pool size and how to achieve so efficiently with {\sys}.
{\sys} complements these single-instance model serving systems with fast autoscaling mechanisms:
we build upon them for fast model serving on a single instance,
and additionally provide ultra-fast scaling when the system needs to change
the number of serving instances.

\stitle{Dynamic scaling serving instances. }
Dynamically scaling serving instances is challenging, mainly because the size of model weights is huge
and still increasing,
so loading them to the accelerators (data plane) is time-consuming.
Some existing works accelerate the loading~\cite{pipeswitch, DBLP:conf/eurosys/JeongBA23, DBLP:conf/asplos/MiaoSDXL0J24, DBLP:conf/osdi/SunHZXZL024}:
For example, both PipeSwitch~\cite{pipeswitch} and DeepPlan~\cite{DBLP:conf/eurosys/JeongBA23} leverage the layer-by-layer character of models
to overlap the inference and parameter loading to hide the loading cost. 
They only focus on host-to-device loading and such overlap is not live especially when the models are large.
SpotServe~\cite{DBLP:conf/asplos/MiaoSDXL0J24} and Llumnix~\cite{DBLP:conf/osdi/SunHZXZL024}
realize live migration but migration cannot fully unleash the computing
capabilities of both instances.
{\sys} provides a new mechanism to scale serving instances lively during parameter loading,
resulting in throughput increase even with unfinished loading,
which we have shown critical in reducing latencies under bursty workloads.

A concurrent work $\lambda$Scale~\cite{yu2025lambdascale} also focuses on using network to accelerate model autoscaling. 
The key difference is that $\lambda$Scale scatters the parameters scaling on multiple instances
to reduce the time for scaling at the cost of decreased serving throughput,
while {\sys} seamlessly scales full parameters on all instances with a similar speed,
yet does not sacrifice the throughput thanks to our multicast-chain-based scaling.
Moreover, during scaling {\sys} has a gradually increasing throughput thanks to our live scaling
while $\lambda$Scale is still a stop-the-world approach.

\stitle{Accelerating coldstart in serverless computing. }
Accelerating model scaling builds upon coldstart acceleration in
serverless computing~\cite{sock,sand,FAASM,DBLP:conf/asplos/DuYXZYQWC20,DBLP:conf/asplos/Ustiugov0KBG21,DBLP:conf/eurosys/SaxenaJSKA22,DBLP:conf/usenix/WangCTWYLDC21},
which focuses on starting general-purpose computing instances like containers. 
We built upon these works, e.g., for accelerating container startup time,
yet designed efficient network-based live autoscale tailored for model scaling
with the domain-specific knowledge of model serving.

\section{Conclusion and Future Work}
\label{sec:concl}

\noindent
Autoscaling is the key to achieving both high goodput and hardware utilization
in model as a service systems,
but the performance overhead introduced by current slow and stop-the-world autoscaling
significantly limits its effectiveness.
In this paper, we first show that the data plane of model autoscaling
can be made fast with less than $O(1)$ caching by leveraging
network-based model-aware multicast.
We next show that the data plane can be made live through model-aware remote execution.
Equipped with these two techniques,
our system {\sys} has at most 94.1\,\% better performance
and 19.46\,\% better resource utilization compared to state-of-the-art serving systems
with and without autoscaling, respectively.
We believe our work
demonstrates the potential and practicability of autoscaling-empowered model
as a service systems.

While fast and live autoscaling of {\sys} takes a key step toward
modern elastic serving systems,
several challenges remain.
First,
during our investigations, we found
scaling policies---determining when and how to scale---may also impact system efficiency.
The policy depends heavily on workload characteristics,
which we leave as future work.
Second, {\sys} currently focuses on instance-level scaling,
whereas modern models can scale by changing the parallel configuration within an instance,
e.g., scaling experts in mixture-of-experts (MoE) models.
While {\sys} in principle works for such a setup,
we leave the detailed exploration in the future work.

\section*{Acknowledgment}
\noindent
We would like to thank OSDI reviewers and our shepherd for
their insightful feedback. 
We sincerely thank Qinwei Yang, Xinhao Luo, and Mingcong Han
for giving us feedback on 
GPU communication library, kernel, driver and runtime.
We also thank Xiating Xie, Chenhan Wang, Sundi Guan for refining the presentation of this paper.
We thank Alibaba Tongyi Lab for providing the testbed during the early stage of this work.
This work was supported in part by
the National Key Research \& Development Program of China (No. 2022YFB4500700),
the Fundamental Research Funds for the Central Universities,
the National Natural Science Foundation of China (No. 62202291, 62272291),
as well as a research grant from Huawei Cloud.

\balance

\small{
\bibliographystyle{acm}
\bibliography{fastllmstart}
}

\twocolumn
\appendix
\section{Appendix}
\label{sec:appendix}

\begin{table*}[!ht]
	\centering
	\begin{tabular}{lllllll}
		\toprule
		\textbf{Instance type}                    & \textbf{Accelerators} & \makecell{\textbf{Local SSD}\\\textbf{BW/GPU}} & \makecell{\textbf{Remote SSD}\\\textbf{BW/GPU}} & \makecell{\textbf{Network}\\\textbf{BW/GPU}} & \textbf{Has NVLink} & \textbf{Price} \\ \midrule
		a2-ultragpu-8g \cite{googleinstance}      & 8 x A100(80\,GB)      & 2.58\,Gbps                & 0.29\,Gbps                 & 12.5\,Gbps              & \ding{51}           & 40.44 USD/h    \\
		p4d.24xlarge\cite{awsinstance}            & 8 x A100(40\,GB)      & 2.31\,Gbps                & -                          & 100\,Gbps               & \ding{51}           & 45.039 USD/h   \\
		ml.hpcpni2.28xlarge\cite{volcanoinstance} & 8 x A100(80\,GB)      & 4\,Gbps                   & -                          & 100\,Gbps               & \ding{55}           & 48.23 USD/h    \\
		p4de.24xlarge\cite{awsinstance}           & 8 x A100(80\,GB)      & 2.31\,Gbps                & -                          & 100\,Gbps               & \ding{51}           & 56.328 USD/h   \\
		a3-highgpu-8g\cite{googleinstance}        & 8 x H100              & 6.09\,Gbps                & 0.97\,Gbps                 & 100\,Gbps               & \ding{51}           & 88.25 USD/h    \\
		a3-megagpu-8g\cite{googleinstance}        & 8 x H100              & 6.09\,Gbps                & 0.97\,Gbps                 & 200\,Gbps               & \ding{51}           & Unavailable    \\
		p5.48xlarge \cite{awsinstance}            & 8 x H100              & 9.8\,Gbps                 & -                          & 400\,Gbps               & \ding{51}           & Unavailable    \\ \bottomrule
	\end{tabular} \\[10pt]

    \begin{minipage}{1\linewidth}
        \caption{\small{\textit{
		A survey of {\maas} hardware configurations from GPU vendors.
        }}}
    \label{tab:vendor-hardware}
    \end{minipage} 
\end{table*}

\subsection{Notable implementation details}
\label{sec:appendix-implementation}

\nospacestitle{Network library }
We implement a communication library
that abstracts both NVLink and RDMA to holistically transfer parameters,
similar to DeepEP~\cite{deepep}.
During our implementation, we found 
establishing communication group between machines is slow (e.g., 100\,ms) 
when using off-the-shelf group communicators (e.g., NCCL~\cite{nccl}),
which significantly limit the effectiveness of network-based scaling.
Fortunately, we found that our plan only requires P2P communication between each pair of nodes. 
Therefore, we pre-create a connection pool that supports full-mesh connections on each.
While the compute network (RDMA) has potential scalability issue~\cite{fasst}, 
it only occurs when transferring small payloads and can be addressed using advanced RDMA 
transport like DCT~\cite{krcore}. 

\stitle{Native serving engine with CUDA context pool. }
Before execution, a CUDA context with loaded kernels (cuModule) must be created on GPU.
Creating such a CUDA context takes about 500\,ms, and is non-negligible in serving instance
autoscaling. To mitigate such an overhead, {\sys} preserves a small CUDA context pool with
pre-loaded kernels and transfers parameters to GPU within one of the existing CUDA contexts,
similar to an existing work~\cite{DBLP:journals/corr/abs-2405-12079}.
Furthermore, {\sys} is built using C++ and native CUDA APIs, eliminating the overhead of initializing
PyTorch (e.g. \texttt{dlopen}).

\stitle{Fault tolerance. }
When machine failures occur, 
we will autoscale new instances using our scaling mechanism. 
One problem is that cached parameters on the failed machine are lost, 
so we need to redistribute these parameters to other machines to maintain 
our global parameter pool invariant. 
For other components in the system like 
scheduler or monitor failures, we follow the same procedure
in existing work for recovery~\cite{DBLP:conf/usenix/Romero0YK21,DBLP:journals/corr/abs-2401-14351}.

\subsection{Hardware configurations for {\maas}}
\label{sec:appendix-hardware}

\noindent
Table~\ref{tab:vendor-hardware} lists the hardware configurations backed by typical 
{\maas} systems. 

\clearpage

\end{document}